\documentclass{JHEP3}

\title{From  Poincare  Invariance to Gauge Theories:   Yang-Mills  and General Relativity }
\author{B. Sazdovi\'c \\
        Institute of Physics, P.O.Box 57, 11001 Belgrade, Serbia \\
        E-mail: \email{sazdovic@phy.bg.ac.rs} }

\abstract{

This article is founded on two fundamental principles: the principle field equations introduced in Refs.\cite{S, S1, S2} and the Fock-Ivanenko covariant derivatives \cite{FI, F}. The former yields the equations of motion for free fields of arbitrary spin and helicity. In the massless case, it also dictates that Lorentz transformations for tensor fields acquire an additional term, which takes the form of a gauge transformation  \cite{W, S1}.

The latter principle, the Fock-Ivanenko derivative, introduces interactions based on the intrinsic and Poincare  groups. This framework allows us to recover a complete Yang-Mills theory, as well as  general relativity in the connection-based formulations of Palatini and Ashtekar, both of which are theories with local gauge symmetries.

While the standard approach begins with the symmetries of a matter action, we will instead derive dynamics directly from  Poincare  invariance. This perspective reveals that   for free fields, Lorentz invariance induces the gauge symmetry of massless tensors. A proper definition of these gauge transformations, in turn, requires the covariant derivatives provided by the Fock-Ivanenko approach.

Considering  matter fields,  we derive the interacting Dirac equation in the presence of Yang-Mills and gravitational fields from its free counterpart.

 }


\usepackage{amsmath}

\begin{document}

\maketitle


\section{Introduction  }
\setcounter{equation}{0}

The Poincare  group plays a fundamental role in contemporary physics. As discovered by Weinberg \cite{W} and further  confirmed in \cite{S,S1, S2}, Poincare  symmetry completely defines the equations of motion in free field theory for particles of arbitrary spin and helicity.   Moreover, it   provides  the origin of local gauge invariance.   On the other hand, localizing the Poincare  symmetry leads to the theory of general relativity \cite{Ut, Ki, Sc, MB}.

There have been similar attempts before to consider arbitrary spin and helicity. An interesting approach involves constraining the spin to eliminate negative-metric components. This constraint,   together   with the Klein-Gordon equation, selects an irreducible unitary representation of the Poincare  group. For further details and references, see \cite{Siegel}.

In this work, we adopt Wigner's definition of particles as irreducible representations of the Poincare group \cite{Wi, BW}. In our previous research \cite{S, S1, S2},  we derived the equations of motion for free massive particles of arbitrary spin and massless particles of arbitrary helicity.   We note that these equations are essentially eigenvalue equations for the Casimir operators of the Poincare group.

In certain important massless cases, the components of the Pauli-Lubanski vector do not commute. Consequently, no common eigenvectors exist for all components. This feature has an unexpected consequence: the Lorentz transformations for vector and symmetric second rank tensor fields acquire an additional term that takes the form of a gauge transformation \cite{W, S1}.

In this paper,   we generalize the framework for free particles to the case of interacting particles. This approach leads naturally to Yang-Mills theory and general relativity as theories with local gauge symmetries. Note that, unlike the standard approach which begins with the symmetries of the matter action, we will start from the gauge symmetry of tensor fields, which is obtained from the requirement of Lorentz invariance.

The key element is the gauge transformation of the tensor field, which is given by the derivative of the gauge parameters. In Maxwell's theory, we work with a single vector field. Thus, the gauge parameter is a scalar and the local symmetry group is $U(1)$.

When multiple vector fields are present, instead of a simple phase rotation, we must consider symmetries associated with a continuous group $G$. A crucial difference in this case is that the ordinary derivative is not well-defined, as it acts on tensor fields defined on the group manifold. As we will show, the correct definition of the derivative leads to the Fock-Ivanenko covariant derivative \cite{FI, F}, thereby introducing fundamental geometrical  and physical  principles.

The Fock-Ivanenko derivatives introduce new fields, known as connections, which govern parallel transport. Since these connections reflect the structure of the group $G$, they are independent of the specific fields they act upon. They possess a deep physical interpretation as the interaction fields.

Replacing the ordinary derivatives with Fock-Ivanenko covariant derivatives, not only introduces the connection but also leads to the non-commutativity of the derivatives. In fact, the commutator of two covariant derivatives yields a function, not an operator, known as the field strength. This allows us to construct gauge and Lorentz invariants, which can be used as Lagrangians for the interacting fields.

Technically, we replace the ordinary derivatives in the action for matter fields obtained in \cite{S, S1, S2} with Fock-Ivanenko covariant derivatives. This procedure introduces interactions between the matter fields and the gauge fields.

In the first part of this paper, we will briefly review elements of the free theory from \cite{S, S1, S2}. We will extend that theory by adding the free theory of a third rank tensor field with helicity $\lambda = 2$, with a particular focus on the gauge transformations of the free fields.
In the second part, we will demonstrate that generalizing this framework to a set of fields leads to complete interacting theories, including Yang-Mills theory and general relativity in the Palatini and Ashtekar formulations.

To describe interactions with matter fields, we will follow \cite{FI, F} and derive the Dirac equation in the presence of Yang-Mills and gravitational fields, starting from the free Dirac equation.


\section{Representations of Casimir  Poincare invariants}
\setcounter{equation}{0}

We will need  representations of  the Poincare algebra  as a first step  in order to find representations of  the Poincare Casimir invariants.

\subsection{Poincare   algebra  }

Lie algebra of Poincare group has a form
\begin{eqnarray}\label{eq:PG}
[ P_a, P_b] = 0  \, , \qquad  [P_c, M_{a b}]  =   - i \Big( \eta_{b c} P_a  - \eta_{a c } P_b  \Big)   \, ,
\end{eqnarray}
\begin{eqnarray}\label{eq:}
[M_{a b}, M_{c d}]  =  i   \Big( \eta_{a d} M_{b c} + \eta_{b c} M_{a d} - \eta_{a c} M_{b d} - \eta_{b d} M_{a c} \Big)   \,  ,
\end{eqnarray}
where $P_a$ are translation generators and
$ M_{a b}   =   L_{a b} +   S_{a b}$  are four dimensional  rotations generators.  They  consist of orbital part
$L_{a b} =    x_a P_b - x_b P_a $  and spin part $S_{a b}$.

\subsection{Poincare  Casimir operators}

Casimir operators  commute  with all  group  generators and  allow us to label the irreducible representations.
Here we will introduce Casimir operators for massive  and massless    Poincare  group and find theirs representations for arbitrary spins  and helicities.

Let us first introduce  Pauli-Lubanski vector
\begin{eqnarray}\label{eq:PLv}
W_a = \frac{1}{2} \varepsilon_{a b c d} M^{b c} P^d     \,   .
\end{eqnarray}
It  does not depend  on orbital part  $L_{a b} $  because
\begin{eqnarray}\label{eq:PLv1L}
 \varepsilon_{a b c d} L^{b c} P^d =   \varepsilon_{a b c d}  (x^b P^c - x^c P^b  )   P^d   =  0  \, .
\end{eqnarray}
Therefore, we can rewrite  Pauli-Lubanski vector   in the form
\begin{eqnarray}\label{eq:PLv1}
W_a = \frac{1}{2} \varepsilon_{a b c d} S^{b c} P^d     \, ,
\end{eqnarray}
where    $S_{a b}$ is spin part of Lorentz generators   $M_{a b}$.

Let us stress that   from Casimir  operator's point of view  just  generators $P_a$  and $S_{a b}$   are relevant. They satisfy relations
\begin{eqnarray}\label{eq:PGS}
[P_a, P_b] = 0  \, , \qquad  [P_c, S_{a b}]  =  0  \, ,
\end{eqnarray}
\begin{eqnarray}\label{eq:PGSS}
[S_{a b} , S_{c d}]  =  i   \Big( \eta_{a d} S_{b c} + \eta_{b c} S_{a d} - \eta_{a c} S_{b d} - \eta_{b d} S_{a c} \Big)   \,  .
\end{eqnarray}
Note that generators  $P_a$ and  $S_{a b}$ commute and so  Casimir  operators   are defined  without problem of order  unambiguity.

\subsubsection{Massive case }

In  the case of  $P^2 > 0$  there are   two Casimir operators
\begin{eqnarray}\label{eq:COP1}
P^2 = P^a P_a \, ,  \qquad     W^2  =  - \frac{P^2}{2}  S_{a b}  S^{a b}  \,   .
\end{eqnarray}

Representations of Poincare group  are labeled by the  eigenvalues of Casimir invariants,   that we can assign to a physical state.  For $P^2 > 0$ eigenvalues of Casimir operators are    mass $m$  and   spin  $s$
\begin{eqnarray}\label{eq:EvCo}
P^2 = m^2 \, ,  \qquad  W^2 =   - m^2   s(s+1) \, .
\end{eqnarray}

\subsubsection{Massless case }

In   massless case   where  $P^2 = 0$,   there are   two Casimir operators for Poincare group   the  helicity  $\lambda$ and   the sign of $P_0$.
The covariant  equation that define helicity   has a form
\begin{eqnarray}\label{eq:COP10}
 W_a   =  \lambda P_a           \,  ,  \qquad  W_a = \frac{1}{2} \varepsilon_{a b c d} S^{b c} P^d     \,   .
\end{eqnarray}
Since   $P^a W_a = 0$ we can conclude    that $P^2 = 0$   is consequence of  (\ref{eq:COP10}).

\subsection{Representation of Casimir operators}

We will use notation $\Psi^A (x) $  for arbitrary field where $A$  is a set of Lorentz vector and spinor indices.

Representations of Poincare group  are labeled by the  eigenvalues of Casimir invariants,   that we can assign to a physical state.
To find  representation of Casimir operators we need representation of Poincare algebra  generators,  momentum  $P_a$  and spin   $ S_{a b}$.
Representation of momenta   $P_a$ is well known from quantum mechanics $(P_a)^A{}_B \to i   \delta^A_B  \partial_a$ and it is  spins independent.

We can  obtain   representation of   spin generators  $ (S_{a b})^A{}_B$  for  arbitrary field   from corresponding expressions with smaller spins and initial expression for fermions.
Starting from infinitesimal Lorentz  transformation
\begin{eqnarray}\label{eq:iLt}
\Psi_\omega^A (x)  = \Psi^A (x) - \frac{i}{2} \omega^{a b}  (S_{a b})^A{}_B  \Psi^B (x)    \, ,
\end{eqnarray}
we can  find    infinitesimal  transformation of product $\Psi_1^A \Psi_2^B$. First,  multiplying expressions   (\ref{eq:iLt})  for  infinitesimal  $ \omega^{a b}$  we obtain
\begin{eqnarray}\label{eq:}
\Psi^A_{1 \omega} (x) \Psi^B_{2 \omega} (x)
 = \Psi^A_1 (x)\Psi^B_2 (x)  - \frac{i}{2} \omega^{a b}   \Big[  (S_{a b})^A{}_C  \Psi^C_1 (x) \Psi^B_2 (x)  +  \Psi^A_1 (x) (S_{a b})^B{}_D  \Psi^D_2 (x)   \Big]   \, .  \qquad
\end{eqnarray}
Second,  by the definition of the Lorentz  transformation  we have
\begin{eqnarray}\label{eq:}
\Psi^A_{1 \omega} (x) \Psi^B_{2 \omega} (x)   =  \Psi^A_1 (x)\Psi^B_2 (x)  - \frac{i}{2} \omega^{a b}  (S_{a b})^{A B}{}_{C D} \Big( \Psi_1^C (x) \Psi_2^D (x) \Big)  \, .
\end{eqnarray}
Comparing these results, we can conclude that  spin generators   $(S_{a b})^A{}_B$ act as derivatives, where their form depend on the fields on the right
\begin{eqnarray}\label{eq:BEder}
 (S_{a b})^{A B}{}_{C D}    =  (S_{a b})^{A}{}_{C} \delta_D^B  +  \delta^A_C (S_{a b})^{B}{}_{D}  \, .
\end{eqnarray}

It is easy to check that
\begin{eqnarray}\label{eq:}
 [ S_{a b},  S_{c d} ]^{A B}{}_{C D}
=    [ S_{a b},  S_{c d} ]^A{}_C     \delta_D^B  +  \delta^A_C  [ S_{a b},  S_{c d} ]^B{}_D     \, ,
\end{eqnarray}
so that $(S_{a b})^{A B}{}_{C D} $ is the solution of  (\ref{eq:PGSS})  because $(S_{a b})^A{}_B  $ is the solution of   this  relation.

The  initial expression for   Dirac spinor is
\begin{eqnarray}\label{eq:gLr21}
(S_{a b})^\alpha{}_\beta   = \frac{i}{4} [\gamma_a, \gamma_b ]^\alpha{}_\beta  \, ,
\end{eqnarray}
and  we  can  find  representations  for all other fields  using recurrence relation (\ref{eq:BEder}). For example, using expression for vectors in terms of spinors  $ V^a (x) =  {\bar \psi} (x)  \gamma^a   \psi (x) $    we can find  the same expression for vector  spin generators as that   obtained by direct calculation
\begin{eqnarray}\label{eq:SOvf}
 (S_{a b})^c{}_d  =  i \Big(    \delta^c_a   \eta_{b d} - \delta^c_b   \eta_{a d}  \Big)     \, .
\end{eqnarray}


\section{Free  field equations }
\setcounter{equation}{0}

In this section, we review essential concepts from Refs. \cite{S, S1, S2}  that will be necessary for the subsequent discussion. It is based on Wigner's definition of  particles  \cite{Wi, BW}   as irreducible representations of    Poincare  group.

Fields are functions defined on Minkowski spacetime that transform under a specific representation of the Poincare  group. They can carry an arbitrary number of vector and spinor indices, which we collectively denote by the index $A$ in the field  $\Psi^A (x)$. To ensure that a field describes a specific particle state, we must impose certain constraints on  $\Psi^A (x)$.

\subsection{Principle field equations}

In field theory, massive particles are defined by their mass and spin. It is therefore natural to describe them using operators whose eigenvalues correspond precisely to these quantities.

Consequently, we postulate the principle field equations for a massive particle of arbitrary spin as a representation of the relations (\ref{eq:EvCo})
\begin{eqnarray}\label{eq:BS}
(P^2)^A{}_B \Psi^B (x) = m \Psi^A (x) \, , \qquad {\cal S}^A{}_B \Psi^B (x) = s (s + 1) \Psi^A (x) \, .
\end{eqnarray}

These are, in fact, Casimir eigenvalue equations. Their Poincare  covariance is guaranteed because the Casimir operators commute with all generators of the Poincare  group. Furthermore, the Casimir operators commute with each other, implying that they possess a complete set of common eigenfunctions. This allows us to impose both eigenvalue equations simultaneously on the same field $\Psi^A (x)$.

In equations (\ref{eq:BS}), the operators $(P^2)^A{}_B$ and ${\cal S}^A{}_B$ are representations of the Casimir operators. Their eigenvalues, $m$ and $s(s+1)$, define the mass $m$ and the  spin  $s$, respectively. The eigenfunctions $\Psi^A (x)$ are therefore  irreducible representations of the Poincare group, which are uniquely labeled by  mass $m$ and spin $s$.

By employing specific representations for the momentum and spin operators, we obtain the following set of differential equations:
\begin{eqnarray}\label{eq:BS1}
\Big( \partial^2 + m^2 \Big) \Psi^A (x) = 0 \, , \qquad {\cal S}^A{}_B \Psi^B (x) = s (s + 1) \Psi^A (x) \, ,
\end{eqnarray}
where
\begin{eqnarray}\label{eq:Oper}
{\cal S}^A{}_B \equiv - \frac{(W^2)^A{}_B}{m^2} = - \frac{1}{m^2} (S^a{}_c)^A{}_C (S^{c b})^C{}_B \partial_a \partial_b + \frac{1}{2} (S{a b})^A{}_C (S^{a b})^C{}_B \, .
\end{eqnarray}

Using expression (\ref{eq:COP10}) and the representations of the momentum and Pauli-Lubanski vector from Ref.\cite{S1}, we postulate the principle field equation for a massless particle of helicity $\lambda$:
\begin{eqnarray}\label{eq:EMmsc2}
(W_a)^A{}_B \Psi^B (x) = \lambda (P_a)^A{}_B \Psi^B (x)  \, .
\end{eqnarray}
As we will show, this single equation encompasses all known free-field equations for massless particles of arbitrary helicity. It also serves as the foundation for local gauge transformations.

\subsection{Principle  field equations  for standard momentum}

So far, we have derived constraints on the fields $\Psi^A (x)$ by treating them as eigenfunctions of the Casimir invariants.

The next step is to construct the projectors onto irreducible representations of the Poincare  group, which correspond to the well-known field equations. To achieve this, it is useful to  shift  to the framework of the standard momentum.

\subsubsection{Massive case}

For massive fields, the Lorentz-invariant functions of the momentum $p^a$ are its square, $p^2 = \eta_{a b} p^a p^b$, and the sign of $p^0$. Since these quantities are fixed, we can choose a standard momentum $k^a$ and express any physical momentum $p^a$ as a Lorentz transformation of $k^a$:
\begin{eqnarray}\label{eq:paLkam}
p^a = L^a{}_b (p) k^b.
\end{eqnarray}

In the massive case where $p^2 = m^2$, we can choose the rest-frame momentum $k^a = (m, 0, 0, 0)$ as the standard momentum. With this choice, the first Casimir constraint (equation \ref{eq:BS1}) is automatically satisfied.

The significant advantage of this approach is that the second differential equation (\ref{eq:BS1}) simplifies to an algebraic equation for the standard momentum, which is much easier to solve. Once the algebraic equation is solved, we can use the relation (\ref{eq:paLkam}) to transform the solutions back to a general $p^a$, and subsequently to the coordinate representation.

For the standard momentum, the spin equation becomes
\begin{eqnarray}\label{eq:BSms}
{\cal S}^A{}_B \Psi^B (k) = s (s + 1) \Psi^A (k),
\end{eqnarray}
where
\begin{eqnarray}\label{eq:Operr1}
{\cal S}^A{}_B = (S^2_i)^A{}B \,  , \qquad    (S_i)^A{}_B = \frac{1}{2} \, \varepsilon_{i j k} (S_{j k})^A{}_B \,  .
\end{eqnarray}

Note that at the standard momentum, the six components of the spin operator $S_{ab}$ are reduced to the three components $S_i$, which are the generators of spatial rotations. These generators form the little group for the massive Poincare  case.

The next step is to solve the eigenvalue problem for the operator ${\cal S}^A{}_B$ in Eq.(\ref{eq:BSms}).
A non-trivial solution for the function $\Psi^A$ exists only if the characteristic polynomial vanishes
\begin{eqnarray}\label{eq:DetSs}
\det \Big( {\cal S}^A{}_B - \lambda \delta^A_B \Big) = 0, \qquad \text{where} \quad \lambda \equiv s ( s + 1).
\end{eqnarray}
In general, this equation may have multiple roots, $\lambda_i$, where $i = 1, 2, \cdots , n$. Each root corresponds to an irreducible representation and the values $s_i$ derived from $\lambda_i$ represent the spins of these representations.

The eigenfunctions $\Psi_i^A$ with definite spin are constructed using projection operators
 \begin{eqnarray}\label{eq:IrrPa}
\Psi_i^A  = ( P_i )^A{}_B \Psi^B   \, ,     \qquad
( P_i )^A{}_B =   \frac{  \Big[  \prod_{j \neq i}^n    \Big(  {\cal S}   - \lambda_j    \Big) \Big]^A{}_B }{  \prod_{j \neq i}^n    \Big(  \lambda_i   - \lambda_j  \Big) }        \,  . \qquad
 i =  \{ 1, 2, \cdots , n  \} \, ,
\end{eqnarray}
In the case of   degeneracy,   a suitable basis must be chosen     in subspace  $\Xi_\lambda $   of the eigenvalue $\lambda$.

The fields $\Psi_i^A(k)$ transform as representations of the Poincare group, but they are not necessarily irreducible. To obtain irreducible representations, fields $\Psi_i^A(k)$ must be further decomposed into sets with specific symmetry properties.

\subsubsection{Massless case}

In the massless case, where    Lorentz invariant functions of momentum vanish,
 $p^2 = 0$, we can choose a standard light-like momentum   $k^a = (1, 0, 0, - 1  )$.  Any  momentum $p^a$   can then be expressed as a Lorentz transformation of this standard momentum
\begin{eqnarray}\label{eq:paLka}
p^a =  L^a{}_b  (p)  k^b \, .
\end{eqnarray}

This choice simplifies  differential equations, as they become algebraic equations.   After solving these algebraic equations, the full momentum-dependent solutions can be recovered via the inverse Lorentz transformation.

Applying this to the equations of motion, we obtain the following conditions for the field   $\Psi^A$   at the standard momentum $ k^a$
\begin{eqnarray}\label{eq:SGcom}
   (S_{1 2})^A{}_B  \Psi^B (k)   =  \lambda    \Psi^A (k)         \, ,   \nonumber \\
 (\Pi_1 )^A{}_B \Psi^B (k) = 0    \, , \qquad     (\Pi_2 )^A{}_B \Psi^B (k) = 0    \,  .
\end{eqnarray}
Here, the operators    $(S_{1 2})^A{}_B$,  $ (\Pi_1 )^A{}_B$ and $ (\Pi_2 )^A{}_B$   defined as
\begin{eqnarray}\label{eq:PLLg10}
W_0 =   W_3 =   S_{1 2 } \, ,  \qquad
W_1  \equiv    \Pi_2   =   S_{0 2} - S_{3 2}   \, , \qquad
- W_2    \equiv   \Pi_1   =  S_{0 1} -    S_{3 1}   \,  ,
\end{eqnarray}
 are the representations of the generators of the little group. Helicity $\lambda$ is the eigenvalue of the  rotation generator       $(S_{1 2})^A{}_B$.

As is well known, for massless fields, the little group of the Poincare group is $E (2)$, the group of translations and rotations in a two-dimensional Euclidean plane. Its generators, $S_{1 2}, \Pi_1$  and $\Pi_2$,  satisfy the commutation relations:
\begin{eqnarray}\label{eq:}
[\Pi_1, \Pi_2] =  0     \,  ,   \qquad    [S_{1 2}, \Pi_1] = i \Pi_2    \,  ,   \qquad    [S_{1 2}, \Pi_2] = - i \Pi_1        \,  .
\end{eqnarray}

\subsection{Spectrum of   helicities   and   eigenfunctions}

Solution of  eigenproblem  for  component   $(S_{1 2})^A{}_B$   produces a spectrum   of helicities $\lambda_i$  $(i = 1, 2, \cdots , n)$.  Then we can construct
projection operators   $( P_i )^A{}_B$   and   eigenfunctions  $\Psi_i^A$ corresponding to these eigenvalues.

In order   that $(S_{1 2})^A{}_B$  equation  in (\ref{eq:SGcom})   has nontrivial solutions we must require  that  corresponding  characteristic   polynomial vanishes
\begin{eqnarray}\label{eq:DetS1}
\det \Big(   (S_{1 2})^A{}_B - \delta^A_B  \lambda   \Big)  =  0   \, .
\end{eqnarray}
The zeros of  characteristic   polynomial are  eigenvalues $\lambda_i$.

Then we can construct   projection operators  $(P_i)^A{}_B$ on subspaces  $\Xi_i$  with dimension $d_i = dim \Xi_i$  corresponding to  helicity $\lambda_i$
 \begin{eqnarray}\label{eq:PiAB}
( P_i )^A{}_B
=   \frac{  \Big[  \prod_{j \neq i}^n    \Big(  S_{1 2}   - \lambda_j  \delta  \Big) \Big]^A{}_B }{  \prod_{j \neq i}^n    \Big(  \lambda_i   - \lambda_j  \Big) }          \, ,
\end{eqnarray}
and obtain  appropriate    eigenfunctions
 \begin{eqnarray}\label{eq:IrrPa1}
\Psi_i^A  = ( P_i )^A{}_B \Psi^B       \, .
\end{eqnarray}

\subsection{In massless case Lorentz transformations induce  gauge transformations  }

The solution to the first equation (\ref{eq:SGcom}) yields a complete set of eigenfunctions. What are then the consequences of the two additional equations?

For the standard momentum defined by equations (\ref{eq:SGcom}), the field $\Psi^A (k)$ should be annihilated by both operators $\Pi_1$ and $\Pi_2$. However, explicit calculation demonstrates that this does not occur in physically relevant cases. The origin of this issue lies in the non-commutativity of the Pauli-Lubanski vectors, $[W_a, W_b] \ne 0$.

Since the equations (\ref{eq:SGcom}) are Lorentz invariant, any violation of these conditions would also violate Lorentz invariance. To quantify this violation, we introduce the expression
\begin{eqnarray}\label{eq:Txyvg}
\delta \Psi_i^A (\varepsilon_1, \varepsilon_2 ) (x) = i \Big( \varepsilon_1 \Pi_1 + \varepsilon_2 \Pi_2 \Big)^A{}_B \Psi_i^B (x) \, . \qquad
\end{eqnarray}

The equations of motion must be Lorentz invariant and, consequently, cannot depend on the variation $\delta \Psi_i^A (\varepsilon_1, \varepsilon_2 ) (x)$. This implies that $\delta \Psi_i^A (\varepsilon_1, \varepsilon_2 ) (x)$ must represent a gauge transformation of the field $\Psi_i^B (x)$. Therefore, as Weinberg noted, Lorentz transformations induce gauge transformations in the massless case \cite{W}.

We will demonstrate that the principle field equations for specific spins and helicities coincide with well-known   free field equations. As examples, we derive the massive Dirac equation for spin $s = \frac{1}{2}$ and the equations for massless vector and tensor fields with helicities $\lambda = 1$ and $\lambda = 2$.


\section{ The massive Dirac  field   }
\setcounter{equation}{0}

For a Dirac field $\Psi^A (x) \to \psi^\alpha (x)$, the spin operator representation is
\begin{eqnarray}\label{eq:gLr2f}
(S_{a b})^A{}_B   \to  (S_{a b})^\alpha{}_\beta  =  \frac{i}{4} [\gamma_a, \gamma_b ]^\alpha{}_\beta     \, .
\end{eqnarray}
This leads to the following expressions
\begin{eqnarray}\label{eq:gLr2f1}
S_i   = \frac{1}{2} \, \varepsilon_{i j k} S_{j k}    = \frac{i}{4}  \varepsilon_{i j k}  \gamma_j  \gamma_k   \, ,  \qquad
{\cal S}^\alpha{}_\beta  =   [ (S_i)^2 ]^\alpha{}_\beta   =  \frac{3}{4 }  \delta^\alpha_\beta        \, .
\end{eqnarray}
The spin equation (\ref{eq:BSms}) then becomes
\begin{eqnarray}\label{eq:BSs1d}
  {\cal S}^\alpha{}_\beta      \psi^\beta  (k) =  \lambda   \psi^\alpha  (k)   \, .  \qquad   \lambda  \equiv  s (s + 1)
\end{eqnarray}
Given that  $ {\cal S}^\alpha{}_\beta$     is diagonal, we compute the determinant
\begin{eqnarray}\label{eq:}
\det (  {\cal S} - \lambda  )^\alpha{}_\beta  =   \Big( \lambda -  \frac{3}{4}  \Big)^4 =     0             \,  .
\end{eqnarray}
The solution $\lambda = \frac{3}{4}$ corresponds to a spin of $ s = \frac{1}{2}$. The exponent in the determinant indicates a four-dimensional representation, meaning the field $\psi^\alpha (k) $ possesses four degrees of freedom. In this case, the only projection operator is the trivial one, $P^\alpha{}_\beta = \delta^\alpha_\beta$.

Using equation (\ref{eq:paLka}), we can boost from the standard momentum $k^a$ to an arbitrary momentum $p^a$ and then transition to coordinate space. This procedure yields the standard Klein-Gordon equation for all components  $(\partial^2 + m^2  )  \psi^\alpha  (x) =0 $.

This equation can be linearized into the form
\begin{eqnarray}\label{eq:DiE}
(i\gamma^a \partial_a + m ) \psi^\alpha (x) = 0  \, ,
\end{eqnarray}
where $\gamma^a$ are constant matrices. This is recognized as the Dirac equation. Indeed, applying the operator twice recovers the Klein-Gordon equation provided the matrices satisfy the Clifford algebra
\begin{eqnarray}\label{eq:Gam}
\{\gamma^a, \gamma^b\}  = 2 \eta^{a b} \, .
\end{eqnarray}

Therefore, for a field of spin $s = \frac{1}{2}$, the   principle   equations lead directly to the Dirac equation (\ref{eq:DiE}) and its associated gamma-matrix condition (\ref{eq:Gam}).


\section{Massless vector field with helicity $\lambda = 1$ and the Maxwell equations}
\setcounter{equation}{0}

Let us consider the case of a massless vector field. The study of massless tensor fields of arbitrary rank can  be reduced to this foundational case. Furthermore, the massless vector field provides the framework for describing the electromagnetic field.

\subsection{Eigenvalues and projection operators}

For a vector field, where the indices   $A, B $   become Lorentz vector indices    $a, b $  and the field   $\Psi^A $  becomes     $V^a$, the spin   generator in the vector representation is given by
\begin{eqnarray}\label{eq:SOvf1}
(S_{a b})^A{}_B    \to  \Big( S_{a b}  \Big)^c{}_d = i \Big( \delta^c_a \, \eta_{b d}  - \delta^c_b \, \eta_{a d}  \Big)   \,  ,         \qquad    \Rightarrow  \qquad
( S_{1 2 } )^a{}_b  =    i \Big(    \delta^a_1   \eta_{2 b} - \delta^a_2   \eta_{1 b}  \Big)     \, .
\end{eqnarray}

The eigenvalues  $\lambda$ are determined by the consistency condition that the characteristic polynomial  vanish
\begin{eqnarray}\label{eq:CharPoly}
 \det \Big( {\cal S}_{1 2} - \lambda   \Big)^a{}_b    =  \lambda^2 (\lambda - 1  ) (\lambda + 1  ) = 0    \, .
\end{eqnarray}
This yields the following solutions for the eigenvalues
\begin{eqnarray}\label{eq:SEIV}
\lambda_0  =  0 \, ,   \qquad   \lambda_{\pm  1}  =  \pm 1 \, .  \qquad
\end{eqnarray}

The eigenspace  $\Xi_0$  corresponding to    $\lambda_0 = 0$   is two-dimensional ($d_0 = dim \Xi_0 = 2$), meaning its eigenvector carries two degrees of freedom. In contrast, the eigenspaces   $d_{\pm 1}  =  dim \Xi_{\pm 1} = 1$   corresponding to  $\lambda_{\pm 1} = 1$   are one-dimensional ($d_{\pm 1}  =  dim \Xi_{\pm 1} = 1$), with each eigenvector carrying a single degree of freedom.

The general expression for the projection operators from Eq. (\ref{eq:PiAB}), applied to a vector field with $n=3$, takes the form
\begin{eqnarray}\label{eq:PiAB1v}
( P_i )^a{}_b  =    \frac{    \prod_{j \neq i}^3    \Big(  (S_{1 2})^a{}_b   - \lambda_j  \delta^a_b  \Big) }{  \prod_{j \neq i}^3    \Big(  \lambda_i    - \lambda_j   \Big) }         \, .
\end{eqnarray}
These operators can be written explicitly as
\begin{eqnarray}\label{eq:}
( P_{ 0} )^a{}_b (k)
=  \delta^a_b    -    (S_{1 2}^2)^a{}_b   =  \delta^a_b    -   \delta^a_\alpha \delta_b^\alpha    \, , \qquad  (\alpha = 1, 2 )     \nonumber \\
( P_{ \pm 1} )^a{}_b   (k)  = \frac{1}{2}    \Big(   (S_{1 2}^2)^a{}_b    \pm   (S_{1 2})^a{}_b \Big)
= \frac{1}{2}    \Big[  \delta^a_\alpha \delta_b^\alpha     \pm   i   \Big(    \delta^a_1   \eta_{2 b} - \delta^a_2   \eta_{1 b}  \Big)    \Big]   \, .
\end{eqnarray}

\subsection{Basics  vectors  }

To facilitate further analysis, we introduce basis vectors within the previously defined eigenspaces. Specifically, we define the vectors   $k^a$  and  $q^a$   in the two-dimensional eigenspace  $\Xi_0 $,   and the vectors $ \breve{p}_\pm^a $ in the one-dimensional eigenspaces    $  \Xi_{\pm 1}$. This provides a complete set of basis vectors
$e^a_i = \{k^a, q^a,  \breve{p}_+^a,  \breve{p}_-^a  \}$  where $ i=\{ 0+, 0-, +1, -1  \} $
defined explicitly as
\begin{eqnarray}\label{eq:kqp}
k^a   = \delta^a_0 -   \delta^a_3   \, ,  \quad
q^a    =  \delta^a_0 +  \delta^a_3  \, ,  \quad
 \breve{p}_\pm^a    =  \delta^a_1  \pm i     \delta^a_2      \, .  \qquad
\end{eqnarray}

It is straightforward to verify that these vectors satisfy the eigenvalue equation
\begin{eqnarray}\label{eq:}
(S_{1 2})^a{}_b e^b_i = \lambda_i e^a_i  \, ,
\end{eqnarray}
where the eigenvalues    $\lambda_i$  are given by (\ref{eq:SEIV}). Consequently, each basis vector $e^a_i$  carries a helicity   $\lambda_i$. In particular, the vectors  $ \breve{p}_\pm^a$  carry helicities $\pm 1$, while both  $k^a$    and  $q^a$     carry helicity    $0$.

If a field   $\Psi^A$  contains $n_+$  factors of $ \breve{p}_+^a$ and $n_-$ factors of  $ \breve{p}_-^a$, its helicity is given by
\begin{eqnarray}\label{eq:Hln}
\lambda = n_+ - n_-  \, .
\end{eqnarray}
Tensors with the highest possible helicity, such as a rank-$n$ tensor with helicities    $\lambda_{\pm n}  = \pm n$,  take the form   $\breve{p}_\pm^{a_1}  \breve{p}_\pm^{a_2} \cdots \breve{p}_\pm^{a_n} \mathbf{T} $.     Therefore, these highest-helicity tensors are one-dimensional (in the space of helicity states) and are symmetric in all their indices.

The projection operators can be expressed in terms of these basis vectors as follows
\begin{eqnarray}\label{eq:PRkqp5}
( P_{ 0 +} )^a{}_b (k)  =  \frac{1}{2}   q^a  k_b  \, , \qquad   ( P_{ 0 -} )^a{}_b (k)  =  \frac{1}{2}   k^a  q_b  \, , \qquad
( P_{ \pm 1} )^a{}_b   (k)  = - \frac{1}{2}   \breve{p}_\pm^a   (\breve{p}_\mp)_b   \, .
\end{eqnarray}
We can confirm that these expressions are indeed projectors by using the inner products of the basis vectors, where the only non-vanishing products are
\begin{eqnarray}\label{eq:SPbv}
k^a q_a = 2  \, ,   \qquad   \breve{p}_\pm^a   (\breve{p}_\mp)_a  =  - 2    \, .
\end{eqnarray}
Note that all basic vectors are light-like.

\subsection{Eigenfunctions of operator $(S_{1 2})^a{}_b$  }

The eigenfunctions of operator $(S_{1 2})^a{}_b$     are   given by  $V_i^a  (k ) =  (P_i)^a{}_b V^b $  or explicitly
\begin{eqnarray}\label{eq:Vfbv}
V_{0 +}^a  (k )  =     \frac{1}{2}   q^a  k_b V^b    \equiv q^a \mathbf{V}_k   \, , \qquad
 V_{0 -}^a  (k ) =     \frac{1}{2}   k^a  q_b V^b  \equiv k^a \mathbf{V}_q   \, ,     \qquad   \nonumber \\
  V_{\pm 1}^a  (k )    =    - \frac{1}{2}    \breve{p}_\pm^a   (\breve{p}_\mp)_b   V^b  \equiv   \breve{p}_\pm^a  \mathbf{V}_\mp             \,  .   \qquad
\end{eqnarray}
It is straightforward to verify   that relation  (\ref{eq:Hln})  is satisfied  as scalar quantities such as   $\mathbf{V}_\mp = (\breve{p}_\mp)_b   V^b$  carry   zero helicity.

The basis vectors define the properties of the eigenvectors. They share the same helicity, gauge transformation rules, and parity.
All eigenfunctions  $V_i^a  (k )$   are  potential representations of the Poincare group.  However, only the gauge-invariant ones constitute true representations of the group.

Under space inversion,  the vectors  $k^a$  and $q^a$ are invariant  while
\begin{eqnarray}\label{eq:}
{\cal P}  \breve{p}_\pm^a    =  -   \breve{p}_\mp^a  \, .
\end{eqnarray}
and consequently
\begin{eqnarray}\label{eq:}
{\cal P}   ( P_{ \pm 1} )^a{}_b   (k)  =    ( P_{ \mp 1} )^a{}_b   (k)     \,  ,     \qquad
{\cal P}    V_{ \pm 1}^a (k)  =    V_{ \mp 1}^a     (k)   \, .
\end{eqnarray}

\subsection{Gauge transformations   of  massless vector  fields }

The gauge transformations for the eigenfunctions of the operator  $(S_{1 2})^a{}_b$  for  massless vector fields
\begin{eqnarray}\label{eq:Txyvgv}
\delta  V_i^a   ( \varepsilon_1, \varepsilon_2 )  (k)
=   i \Big(  \varepsilon_1 \Pi_1  +  \varepsilon_2 \Pi_2  \Big)^a{}_b    V_i^b  (k)   \, , \qquad
\end{eqnarray}
 constitute a particular case of Eq.(\ref{eq:Txyvg}). The transformations for the basic vectors themselves have a simple form
\begin{eqnarray}\label{eq:gtbv}
\delta  k^a   = 0   \, ,  \qquad
\delta  q^a  =       \varepsilon_+   \breve{p}_-^a   +   \varepsilon_-   \breve{p}_+^a       \, ,      \qquad
 \delta  \breve{p}_\pm^a  =   \varepsilon_\pm   k^a    \, .  \qquad
( \varepsilon_\pm =  \varepsilon_1  \pm i    \varepsilon_2    )   \qquad
\end{eqnarray}
The gauge transformations of the eigenfunctions   $V_i^a  (k ) =  (P_i)^a{}_b V^b (k)$    are consequently determined by those of the basic vectors.

It is important to note that only the component  $ V_{0 -}^a  (k )$    is gauge invariant, satisfying $ \delta  V_{0 -}^b  (k ) =  0 $. As a result, only this component constitutes an irreducible representation of the Poincare group. The remaining components,    $ V_{0 +}^a  (k )$  and  $V_{\pm 1}^b  (k )$, are not gauge invariant and transform as follows
\begin{eqnarray}\label{eq:GtrV}
\delta   V_{0 +}^a  (k )  =    \omega_+ \breve{p}_-^a  +      \omega_-  \breve{p}_+^a        \, ,  \qquad
  \omega_\pm   (k )   = \frac{1}{2}  k_a  V^a  \varepsilon_\pm     \, ,   \nonumber \\
\delta    V_{\pm 1}^a  (k )   =    k^a  \Omega_\pm      \, ,     \qquad
\Omega_\pm   (k ) =    - \frac{1}{2}   (\breve{p}_\mp)_b    V^b   \varepsilon_\pm      \, .
\end{eqnarray}

Furthermore,  from  (\ref{eq:gtbv}) we can  deduce  that under space inversion  ${\cal P} \varepsilon_\pm = - \varepsilon_\mp$.  This implies the transformation properties   ${\cal P} \omega_\pm = - \omega_\mp$  and
${\cal P} \Omega_\pm =  \Omega_\mp$.    These results are consistent with the transformations of the eigenfunctions   ${\cal P}  V_{0 +}^a  (k ) =  V_{0 +}^a  (k )$   and  ${\cal P}  V_\pm^a  (k ) =  V_\mp^a  (k )$.

\subsection{Action  for  massless vector  field with helicities  $\lambda = 1$ yields   Maxwell equations }

The electromagnetic interaction is symmetric under spatial inversion. Consequently, rather than working with the two independent components   $ V_{+ 1}^a  (k)$  and $ V_{- 1}^a  (k )$  which individually form incomplete irreducible representations of the  proper  Poincare  group we introduce a single vector field $A^a (k)$. This field constitutes an incomplete irreducible representation of the extended group that includes spatial inversion
\begin{eqnarray}\label{eq:GSbh1v}
 A^a   (k ) =   \alpha  V^a_{+ 1}  (k )  +  \beta  V^a_{- 1}  (k )  \, .  \qquad
\end{eqnarray}
In line with this, we also combine the two gauge parameters   $\Omega_+ (k )$  and  $\Omega_- (k )$, which are related by spatial inversion, into a single parameter   $\Omega (k ) =  \alpha \Omega_+ (k )  +  \beta \Omega_- (k ) $.

The representations  $ V_{+ 1}^a  (k)$  and $ V_{- 1}^a  (k )$  are one-dimensional, meaning each possesses one degree of freedom. The combined field   $A^a (k)$   therefore carries two degrees of freedom and, as will be shown, describes the photon.

To construct a gauge-invariant action in terms of   $ A^a  (k )$, we first promote the standard momentum
$k^a $ to an arbitrary momentum  $p^a$, and then transition to coordinate space fields via the substitution   $p^a = i \partial^a$. For a real field  $A^a (x)$, Eq. (\ref{eq:GSbh1v}) gives
\begin{eqnarray}\label{eq:}
A^a (x) =  \int d^4 p  \Big( A^a (p)  +  A^a (- p)   \Big) e^{- i p x }   \nonumber \\
=  \int d^4 p  \Big[   \alpha  \Big(   V^a_{+ 1}  (p )  +   V^a_{+ 1}  (- p )   \Big)  +   \beta \Big(   V^a_{- 1}  (p )   +   V^a_{- 1}  (- p )      \Big)     \Big] e^{- i p x }     \, .
\end{eqnarray}

Using the second line of Eq.(\ref{eq:GtrV}), the gauge variation of the field is
\begin{eqnarray}\label{eq:}
\delta  A^a (x)
=  \int d^4 p \,   p^a  \Big[   \alpha   \Big(  \Omega_{+ 1}  (p )  -   \Omega_{+ 1}  (- p )   \Big)
+   \beta \Big(   \Omega_{- 1}  (p )   -  \Omega_{- 1}  (- p )      \Big)     \Big] e^{- i p x }             \, ,
\end{eqnarray}
which simplifies to
\begin{eqnarray}\label{eq:GSbh1}
\delta  A^a (x)
=  \partial^a     \Big(  \alpha  \Omega_{+ 1} ( x )  +  \beta \Omega_{- 1}  (x )   \Big)  =  \partial^a  \Omega (x)       \, .
\end{eqnarray}

The action must be constructed exclusively from gauge-invariant quantities. For a vector field  $A_a$, the fundamental gauge-invariant object is the field strength tensor $ F_{a b} = \partial_a A_b - \partial_b A_a $.   Since the action itself must be a Lorentz scalar, it can be built from scalar combinations of  $ F_{a b}$, such as
\begin{eqnarray}\label{eq:AcI0}
I_0 =  - \frac{1}{4} \int d^4 x    F_{a b} F^{a b}   \, .
\end{eqnarray}

Interactions with other fields are incorporated by adding an interaction term
\begin{eqnarray}\label{eq:Inm}
I_{int} \Big( A^a \Big)  =  \int d^4 x   A^a J_a  \, ,
\end{eqnarray}
to the action. Gauge invariance imposes the requirement that $I_{int}$   remains unchanged under a gauge transformation. For an infinitesimal transformation, this means
 $ I_{int} \Big( A^a + \delta  A^a  \Big) =  I_{int} \Big( A^a  \Big)$.

 This condition becomes, after partial integration,  $ \int d^4 x    \Omega  \,   \partial^a   J_a     =     0  $.
Since this must hold for an arbitrary function  $ \Omega (x)$, it follows that the current must be conserved
 $\partial^a J_a  = 0$.

The complete action for electrodynamics is therefore
\begin{eqnarray}\label{eq:}
I  =  I_0  +  I_{int} =  \int d^4 x  \Big(   - \frac{1}{4}   F_{a b} F^{a b}   +   A^a J_a   \Big)    \, .
\end{eqnarray}
Varying this action with respect to the potential  $A^a$ yields the inhomogeneous Maxwell equations.


\section{Massless  fields   with helicity $\lambda = 2$   }
\setcounter{equation}{0}

The previous three sections reviewed necessary material from Refs.\cite{S, S1, S2}. In the following sections, we will derive new equations using the same method.

Our primary objective is to construct a theory of interacting fields. The first step towards this goal is to determine the gauge transformations for free fields using our prescription, which is based on the principle field equations  This will be accomplished in the present section. The subsequent step will be to find the gauge transformations for the interacting fields.

\subsection{Arbitrary rank massless  tensors   }

An $n$-rank tensor field with helicity $\lambda_i$ takes the form \cite{S1}
\begin{eqnarray}\label{we eq:GThl}
 T_{\pm i}^{a_1  a_2 \cdots a_n}   ( k)
=    (P_{\pm i})^{a_1  a_2 \cdots a_n}{}_{b_1  b_2 \cdots b_n}    T^{b_1  b_2 \cdots b_n}  (k)    \,  ,   \qquad
  (i = 0,  1,  2 , \cdots ,  n  )
\end{eqnarray}
where the dimensions of the representations are given by
\begin{eqnarray}\label{eq:Dire}
 d_n  =    1        \, ,   \qquad
 d_{n - 1}    = 2 n        \, , \qquad      \nonumber \\
d_{n - 2}  =   n (2 n - 1)  \, , \qquad
 d_{n - 3}   =   n (n - 1)   \frac{4 n - 2}{3}     \, .   \qquad
\end{eqnarray}

The general expression for the highest-helicity tensor for example, a rank $n$ tensor with helicities
$\lambda_{\pm n} = \pm n$ is given by
\begin{eqnarray}\label{eq:Hhn}
T_{\pm n}^{a_1  a_2 \cdots a_n}  =    (P_{\pm n})^{a_1  a_2 \cdots a_n}{}_{b_1  b_2 \cdots b_n}    T^{b_1  b_2 \cdots b_n}
=       \mathbf{T}_{\mp n}  \, \breve{p}_\pm^{a_1}    \breve{p}_\pm^{a_2}  \cdots \breve{p}_\pm^{a_n}     \, , \qquad
\end{eqnarray}
where
\begin{eqnarray}\label{eq:}
 \mathbf{T}_{\mp n}  =  \frac{(-1)^n}{2^n}  (\breve{p}_\mp)_{b_1}  \,   (\breve{p}_\mp)_{b_2} \, \cdots \,  (\breve{p}_\mp)_{b_n}  \,   T^{b_1  b_2 \cdots b_n}   \, . \qquad
\end{eqnarray}
It is evident that the representation with the highest helicity is totally symmetric and one-dimensional.

\subsection{Gauge transformations for helicity-2 fields}

To describe gravitational interactions, we are particularly interested in fields with helicity $\lambda = \pm 2$.

\subsubsection{Second rank  tensors   with helicity  $\lambda = 2$ }

For a rank-2 tensor ($n=2$)
\begin{eqnarray}\label{eq:}
 T_{\pm  i}^{a b }  (k ) =   ( P_{\pm i} )^{a b }{}_{d e }  \,  T^{d e }  (k)   \,  ,  \qquad
 i = \{ 0,  1,  2     \}              \, ,
\end{eqnarray}
 the projectors onto the helicity states $\lambda = 0, \pm 1, \pm 2$ are given by
\begin{eqnarray}\label{eq:PO02}
(P_0)^{a b}{}_{c d} =  \Big( \pi_0    \pi_0   +   \pi_+  \pi_-  +  \pi_-  \pi_+  \Big)^{a b}{}_{c d}   \, ,  \qquad
(P_{\pm 1})^{a b}{}_{c d}   =   \Big(  \pi_0 \pi_\pm  + \pi_\pm  \pi_0 \Big)^{a b}{}_{c d}   \, , \qquad
    ( P_{\pm 2})^{a b}{}_{c d}     =  \Big( \pi_\pm  \pi_\pm    \Big)^{a b}{}_{c d}   \,   .    \qquad
\end{eqnarray}

The field with the highest helicity, $\lambda = \pm 2$, corresponds to a totally symmetric tensor and forms a one-dimensional representation. In terms of the basis vectors, this field is expressed as
\begin{eqnarray}\label{eq:Hhpm2}
T_{\pm 2}^{a b} (k) = (P_{\pm 2})^{a b }{}_{c d} T^{c d} (k)
=  ( \pi_{ \pm 1} )^a{}_c   ( \pi_{ \pm 1} )^b{}_d  T^{c d}  (k)
=   \breve{p}_\pm^a      \breve{p}_\pm^b      \mathbf{T}_{\mp 2}   \, ,  \qquad
\end{eqnarray}
where we have defined
\begin{eqnarray}\label{eq:}
 \mathbf{T}_{\mp 2} =      \frac{1}{4}   (\breve{p}_\mp)_c   (\breve{p}_\mp)_d    T^{c d} (k)     \, .  \qquad
\end{eqnarray}

By applying the gauge transformation of the basis vectors,  $\delta \breve{p}_\pm^a = k^a \varepsilon_\pm$,  we can derive the corresponding gauge transformation for the helicity-2 field
\begin{eqnarray}\label{eq:}
 \delta  T_{\pm 2}^{a b} (k)   =   k^a  \omega_\pm^b +   k^b  \omega_\pm^a               \, ,  \qquad
 \omega_\pm^a   = \breve{p}_\pm^a     \mathbf{T}_{\mp 2}  \varepsilon_\pm  \, .
\end{eqnarray}

\subsubsection{Third rank tensors with helicity $\lambda = 2$}

For third-rank tensors, where $n = 3$, we obtain
\begin{eqnarray}\label{eq:}
T_{\pm  i}^{a b c}  (k ) =   ( P_{\pm i} )^{a b c}{}_{d e f}  \,  T^{d e f}  (k)  \,  ,  \qquad   i = \{ 0,  1,  2,  3  \}       \, ,
\end{eqnarray}
where the projectors corresponding to helicities $\lambda = 0, \pm 1, \pm 2, \pm 3$ are given by
\begin{eqnarray}\label{eq:Alproj}
( P_0 )^{a b c}{}_{d e f}  =  \Big(  \pi_0  \pi_0 \pi_0 + \pi_0  \pi_+  \pi_- +  \pi_+ \pi_0   \pi_-  + \pi_+  \pi_- \pi_0  + \pi_0 \pi_-  \pi_+  +  \pi_- \pi_0  \pi_+  + \pi_-  \pi_+  \pi_0  \Big)^{a b c}{}_{d e f}
 \, ,    \nonumber \\
  ( P_{\pm 1} )^{a b c}{}_{d e f}   = \Big(  \pi_\pm   \pi_0   \pi_0   +  \pi_0  \pi_\pm     \pi_0  +  \pi_0   \pi_0   \pi_\pm  + \pi_\mp  \pi_\pm  \pi_\pm  + \pi_\pm  \pi_\mp  \pi_\pm  + \pi_\pm  \pi_\pm  \pi_\mp    \Big)^{a b c}{}_{d e f}    \, ,    \nonumber \\
  ( P_{ \pm 2} )^{a b c}{}_{d e f}     =    \Big(  \pi_0    \pi_\pm  \pi_\pm   +  \pi_\pm   \pi_0   \pi_\pm      +     \pi_\pm  \pi_\pm        \pi_0   \Big)^{a b c}{}_{d e f}          \,  ,   \nonumber \\
   ( P_{\pm 3} )^{a b c}{}_{d e f}     =    \Big(   \pi_\pm  \pi_\pm   \pi_\pm    \Big)^{a b c}{}_{d e f}          \,  .    \qquad
\end{eqnarray}
The highest helicity tensor corresponds to $\lambda = 3$. Since we are interested in fields with helicity
$\lambda = 2$, we will use the component associated with the projector  $ ( P_{ \pm 2} )^{a b c}{}_{d e f} $.

To study irreducible representations, we must distinguish tensors with different symmetry properties. We choose a tensor that is antisymmetric in its first two indices
\begin{eqnarray}\label{eq:Tafti}
T_{\pm  i}^{[a b] c}  (k ) =   ( P_{\pm i} )^{a b c}{}_{d e f} \,T^{[d e] f}  (k)    \, .
\end{eqnarray}

This choice is motivated by the fact that, among the three indices, one must correspond to the general connection, while the other two are contracted with the indices of a Poincare  group generator. The only Poincare  generator with two indices is the four-dimensional rotation generator $M_{a b}$, which is antisymmetric in $a$ and $b$. Therefore, we require a tensor that is antisymmetric in two indices, as in equation   (\ref{eq:Tafti}).

To    separate the unwanted  gauge transformation, we retain only the gauge-invariant component of $\pi_0$. Thus, replacing $\pi_0 \to \pi_{0-}$, the projector $(P_{\pm 2})^{abc}{}_{def}$ takes the form
\begin{eqnarray}\label{eq:seagt}
 ( P_{ \pm 2} )^{a b c}{}_{d e f}     =    \Big(  \pi_{ 0 -}    \pi_\pm  \pi_\pm   +  \pi_\pm   \pi_{ 0 -}   \pi_\pm      +     \pi_\pm  \pi_\pm   \pi_{ 0 -}   \Big)^{a b c}{}_{d e f}   \,  .    \qquad
\end{eqnarray}

Using projection operators expressed in terms of basic vectors
\begin{eqnarray}\label{eq:PRkqp51}
( \pi_{ 0 +} )^a{}_b (k)  =  \frac{1}{2}   q^a  k_b  \, , \qquad   ( \pi_{ 0 -} )^a{}_b (k)  =  \frac{1}{2}   k^a  q_b  \, , \qquad
( \pi_{ \pm 1} )^a{}_b   (k)  = - \frac{1}{2}   \breve{p}_\pm^a   (\breve{p}_\mp)_b   \, .
\end{eqnarray}
we   obtain
\begin{eqnarray}\label{eq:tasfi}
 T_{\pm 2}^{[a b] c}  =     \Big(   k^a \breve{p}_\pm^b   -  \breve{p}_\pm^a   k^b    \Big)      \breve{p}_\pm^c  \mathbf{T}_{\pm 3}   \, ,  \qquad
\end{eqnarray}
where we have introduced
\begin{eqnarray}\label{eq:}
\mathbf{T}_{\pm 3}   =    \frac{1}{16} \Big( q_d    \,  (\breve{p}_\mp)_e -  (\breve{p}_\mp)_d  q_e  \Big)    (\breve{p}_\mp)_f     \,  T^{d e f }   \, .  \qquad
\end{eqnarray}

Using the gauge transformations of the basic vectors, $\delta k^a = 0$  and
$\delta \breve{p}_\pm^a = k^a  \varepsilon_\pm$, we can show that the combination
$ k^a \breve{p}_\pm^b   -  \breve{p}_\pm^a   k^b $ is gauge-invariant. From this, we can derive the gauge transformation of the field  $T_{\pm 2}^{[a b] c}  $
\begin{eqnarray}\label{eq:}
 \delta  T_{\pm 2}^{[a b] c}  =   k^c  \omega_\pm^{a b}                \, ,  \qquad
 \omega_\pm^{a b} = \Big(  k^a \breve{p}_\pm^b -  \breve{p}_\pm^a    k^b \Big)
 \mathbf{T}_{\pm 3}   \varepsilon_\pm \, .
\end{eqnarray}

It is useful to specify the dimension of this representation. For the case of  $n = 3$   and $\lambda = 2$ , the dimension, in accordance with Eq. (\ref{eq:Dire}), is $2 n = 6$.
This is consistent with Eq. (\ref{eq:Alproj}), where  $\pi_0$  projects onto a two-dimensional space and
 $\pi_\pm$  onto one-dimensional spaces.
Replacing   $\pi_0 \to \pi_{0-}$    yields the projector in Eq.(\ref{eq:seagt}), which projects onto a three-dimensional space because $\pi_{0 -}$ is one-dimensional.
 Antisymmetrization then eliminates the third term from Eq.(\ref{eq:seagt}), projecting onto a two-dimensional space.   The antisymmetric part of this two-dimensional space is one-dimensional, which is the final dimension of the tensor in Eq. (\ref{eq:tasfi}).

\subsubsection{Higher  rank tensors   $n \ge 4$   with helicity    $\lambda = 2$}

In general, an arbitrary rank-$n$  tensor (for  $n \ge 2$) can be projected onto a field with helicity
$\lambda   =  2$.  So why are we specifically excluding tensors of rank     $n \ge 4$  from this consideration?

As previously explained, one index must be reserved for the general connection. The remaining indices must be contracted with the indices of the Poincare  group generators. There are two such generators: the four-dimensional translation generator $P_a$  and the four-dimensional rotation generator  $M_{a b}$.

While it is possible to contract an arbitrary number of indices with these generators, the resulting expression would be non-linear in the Poincare  generators. Since we require an expression that is linear in the Poincare generators, only two possibilities remain:

Contraction with  $P_a$: This requires a field with two indices--one index $a$  for the translation generator and one for the general connection.

Contraction with  $M_{a b}$: This requires a field with three indices two indices  $(a b)$   for the antisymmetric rotation generator and one for the general connection. Consequently, this field must itself be antisymmetric in the indices  $a$  and  $b$.

\section{ General expression   for connections  }
\setcounter{equation}{0}

This section describes theories with helicities $\lambda = 1$ and $\lambda = 2$, corresponding to Yang-Mills theory and general relativity.

The  gauge field for   vector  theory, with helicity $\lambda = 1$, is given by the expression in \eqref{eq:Vfbv}.
\begin{eqnarray}\label{eq:Vfbv1}
 V_{\pm 1}^a  (k )  =   ( \pi_{ \pm 1} )^a{}_b V^b
 =    - \frac{1}{2}    \breve{p}_\pm^a   (\breve{p}_\mp)_b   V^b
 \equiv   \breve{p}_\pm^a  \mathbf{V}_\mp       \, ,
 \qquad   \mathbf{V}_\mp  \equiv  - \frac{1}{2}     (\breve{p}_\mp)_b   V^b     \,  .   \qquad
\end{eqnarray}

To describe gravity, we require fields with helicity $\lambda = 2$. There are two possibilities. The first is a totally symmetric second-rank tensor, which carries the highest helicity state $n=2$. In terms of the basic vectors, this is represented by \eqref{eq:Hhpm2}
\begin{eqnarray}\label{eq:Hhpm21}
T_{\pm 2}^{a b} (k) =   (P_{\pm 2})^{a b }{}_{c d}   T^{c d} (k) =     ( \pi_{ \pm 1} )^a{}_c   ( \pi_{ \pm 1} )^b{}_d  T^{c d}  (k)
=    \breve{p}_\pm^a      \breve{p}_\pm^b    \mathbf{T}_{\mp 2}    \,   .   \qquad
\end{eqnarray}
The second case is a third-rank tensor given by \eqref{eq:tasfi}
\begin{eqnarray}\label{eq:}
 T_{\pm 2}^{[a b] c}  =     \Big(   k^a \breve{p}_\pm^b   -  \breve{p}_\pm^a   k^b    \Big)      \breve{p}_\pm^c  \mathbf{T}_{\pm 3}    \,   .  \qquad
\end{eqnarray}

The helicity $\lambda = n_+ - n_-$ is determined by the number of vectors     $\breve{p}_\pm^a $.   Consequently, the vector field is linear in     $\breve{p}_\pm^a $, while both cases describing general relativity are bilinear in $\breve{p}_\pm^a$.
Since all fields are one-dimensional, each one carries a single degree of freedom.

\subsection{Unified expression and  gauge transformation for the  connection}

When dealing with multiple field components, it is useful to relate these fields to their corresponding group generators. For a tensor field with a set of indices $R$, we isolate one vector index $a$ (which will later become a coordinate index $a \to \mu$). The remaining set of indices, $I$, so that $R = \{a, I \}$, corresponds to the indices of the appropriate generators.

 For a vector field (Yang-Mills), the index $I$ enumerates the generators of an internal group (e.g., $SU(N)$). Thus, we take
\begin{eqnarray}\label{eq:VfnAg}
  ( \mathbf{ A}_{\pm 1}^a)^A{}_B (k)  =    (V_{\pm 1})^{a I}  (t_I)^A{}_B
  =  \breve{p}_\pm^a  \mathbf{V}_\mp^I  (t_I)^A{}_B    \, ,
\end{eqnarray}
where $(t_I)^A{}_B$ are the group generators and the field   $  ({ A}_{\pm 1}^a)^A{}_B (k)$    is  group  valued
connection.

 For a second-rank tensor (gravity),    $R =\{ a, b \}$ and  $I = b$. Since the momentum is a Poincare generator with one vector index, we take
\begin{eqnarray}\label{eq:}
 ( {}_2   \mathbf{ A}_{\pm 2}^a )^A{}_B  (k) =    (T_{\pm 2})^{a b}    (P_b)^A{}_B
 =     \breve{p}_\pm^a      \breve{p}_\pm^b  \mathbf{T}_{\mp 2}  (P_b)^A{}_B    \,   .   \qquad
\end{eqnarray}

 For a third-rank tensor with helicity $2$, we have $R = [a b] c$ and $I = [a b]$. As the generator of four-dimensional rotations is the antisymmetric   in  two vector indices, we take
\begin{eqnarray}\label{eq:}
 ( {}_3  \mathbf{ A}_{\pm 2}^c )^A{}_B (k)  =    (T_{\pm 2})^{[a b] c}    (M_{a b})^A{}_B
 =    \Big(   k^a \breve{p}_\pm^b   -  \breve{p}_\pm^a   k^b    \Big)
 \breve{p}_\pm^c  \mathbf{T}_{\mp 3}    (M_{a b})^A{}_B      \, .  \qquad
\end{eqnarray}

These last three equations can be written in a unified form
\begin{eqnarray}\label{eq:GFbv}
   ( \mathbf{ A}_\pm^a)^A{}_B  (k)  =     \breve{p}_\pm^a  (   \omega_\pm)^A{}_B  (k)   \, ,
\end{eqnarray}
where for the vector, second-rank, and third-rank tensor cases, we have respectively
\begin{eqnarray}\label{eq:}
 (\omega_{\pm 1})^A{}_B =  \mathbf{V}_\mp^I  (t_I)^A{}_B    \, ,    \qquad
 ({}_2  \omega_{\pm 2})^A{}_B =  \mathbf{T}_{\mp 2}    \breve{p}_\pm^b    (P_b)^A{}_B  \, ,    \qquad
 ({}_3  \omega_{\pm 2})^A{}_B =  \Big(   k^a \breve{p}_\pm^b   -  \breve{p}_\pm^a   k^b    \Big)
 \mathbf{T}_{\mp 3}    (M_{a b})^A{}_B   \, .   \qquad
\end{eqnarray}

Both cases with helicity $2$ can be rewritten as
\begin{eqnarray}\label{eq:}
 ( \mathbf{ A}_{\pm 2}^c )^A{}_B  = (T_{\pm 2})^{c b} (P_b)^A{}_B +(T_{\pm 2})^{[a b] c} (M_{a b})^A{}_B \nonumber \\
=    \breve{p}_\pm^c \Big[   \breve{p}_\pm^b  \mathbf{T}_{\mp 2}  (P_b)^A{}_B  + \Big( k^a \breve{p}_\pm^b   - \breve{p}_\pm^a  k^b \Big)   \mathbf{T}_{\pm 3}    (M_{a b})^A{}_B \Big]     \,   . \qquad
\end{eqnarray}

Therefore, the most general expression for the connection, linear in the generators of both the $SU(N)$ and Poincare groups, is
\begin{eqnarray}\label{eq:MGclg}
 (  \mathbf{ A}_{\pm 2}^c )^A{}_B  =   (V_{\pm 1})^{c I}  (t_I)^A{}_B
 + (T_{\pm 2})^{c b} (P_b)^A{}_B +(T_{\pm 2})^{[a b] c} (M_{a b})^A{}_B   \,  .  \qquad
\end{eqnarray}

Using the gauge transformation of the basic vectors, $\delta \breve{p} _\pm^a = k^a  \varepsilon_\pm$, from \eqref{eq:GFbv}, we can derive the gauge transformation for connection   in the frame of standard momentum
\begin{eqnarray}\label{eq:GFbvc}
\delta   ( \mathbf{ A}_\pm^a)^A{}_B (k)  =    k^a  (\omega_\pm)^A{}_B  (k)  \varepsilon_\pm
\equiv  k^a  ( { \Omega}_\pm)^A{}_B  (k) \,   ,
\end{eqnarray}
 with  gauge parameter
\begin{eqnarray}\label{eq:Gauge parameter}
( { \Omega}_\pm)^A{}_B  (k)
=   \varepsilon_\pm  \Big[  \mathbf{V}_\mp^I  (t_I)^A{}_B  +    \breve{p}_\pm^b  \mathbf{T}_{\mp 2}  (P_b)^A{}_B  + \Big( k^a \breve{p}_\pm^b   - \breve{p}_\pm^a  k^b \Big)   \mathbf{T}_{\pm 3}    (M_{a b})^A{}_B \Big]     \,   .
\end{eqnarray}

\subsection{Coordinate representation of  fields}

 To construct the coordinate representation, we first boost from the standard momentum frame to an arbitrary frame,  and subsequently to coordinate-dependent fields via the mapping
 $k^a  \to p^a \to    i \partial^a$.
Following the approach used in electrodynamics, we define the real fields
\begin{eqnarray}\label{eq:Coore}
{  A}^A{}_B (x) =  \int d^4 p  \Big(  {  A}^A{}_B    (p)  + { A}^A{}_B  (- p)   \Big) e^{- i p x }   \nonumber \\
=  \int d^4 p  \Big[   \alpha  \Big(  ({ A}_+)^A{}_B     (p )   + ({ A}_+)^A{}_B   (- p )   \Big)  +   \beta \Big(   ({ A}_-)^A{}_B    (p )   +   ( { A}_-)^A{}_B   (- p )      \Big)     \Big] e^{- i p x }     \, ,
\end{eqnarray}
and the real local parameters
\begin{eqnarray}\label{eq:}
{  \Omega}^A{}_B (x) =  \int d^4 p  \Big( {  \Omega}^A{}_B    (p)  +  { \Omega}^A{}_B  (- p)   \Big) e^{- i p x }   \nonumber \\
=  \int d^4 p  \Big[   \alpha  \Big(  ( { \Omega}_+)^A{}_B     (p )   + ( { \Omega}_+)^A{}_B   (- p )   \Big)  +   \beta \Big(   ( { \Omega}_-)^A{}_B    (p )   +   ( { \Omega}_-)^A{}_B   (- p )      \Big)     \Big]
e^{- i p x }     \,  .
\end{eqnarray}

 Consequently, the gauge transformation of Eq. (\ref{eq:GFbvc}) takes the following form in coordinate space
\begin{eqnarray}\label{eq:GFbvcg1}
\delta    (  { A}_a)^A{}_B (x)  =  \partial_a    { \Omega}^A{}_B  (x) \, .
\end{eqnarray}


\section{Consistency of the theory  leads to interaction  theory}
\setcounter{equation}{0}

Previously derived   Dirac equation for matter fields (\ref{eq:DiE})  and   the gauge transformation   (\ref{eq:GFbvcg1}) are central for the transition to an interacting theory. The issue is that the derivatives in these equations are not well-defined.   In the case of   spinor field $\psi^\alpha (x)$ we will require consistency between the parallel transport of spinors and tensors.  On the other hand,
 in  the non-Abelian case   the gauge transformations  act on a tensor ${ \Omega}^A{}_B (x) $  rather than on a scalar function $ \Omega(x)$, as in the  case of a single vector field.

The requirement for a consistent theory meaning one with well-defined expressions for matter fields and local gauge transformations in the non-Abelian case necessarily introduces interactions. The crucial step in transition  from a free to an interacting theory is to replace the ordinary derivative with a properly defined covariant derivative, known in the literature as the Fock-Ivanenko derivative (see Appendix \ref{sec:FID}). This derivative incorporates a general connection $( { A}_\mu (x))^A{}_B $.

\subsection{Well defined  theory  requires  covariant derivatives}

Following the procedure from the principle field equations, we obtained the expression for gauge transformations in the coordinate representation (\ref{eq:GFbvcg1}). However, we are faced with a problem because the ordinary derivatives are not well-defined in this context. Consequently, we must replace the partial derivatives $ \partial_a $ with general Fock-Ivanenko covariant derivatives
\begin{eqnarray}\label{eq:CodegA}
    \partial_a   \Omega^A{}_B   (x)   \to   ({\cal D}_\mu \Omega )^A{}_B   (x)
 =   \partial_ \mu  \Omega^A{}_B (x) + i [A_\mu , \Omega]^A{}_B  \, .
\end{eqnarray}

Thus, instead of (\ref{eq:GFbvcg1}), we should use the following expression for gauge transformations
\begin{eqnarray}\label{eq:GSbh1Dn}
\delta  (A_\mu)^A{}_B (x)  =   ({\cal D}_ \mu  \Omega )^A{}_B   (x)
 =   \partial_\mu   \Omega^A{}_B (x) + i [A_\mu , \Omega]^A{}_B       \, .
\end{eqnarray}

The finite form is
\begin{eqnarray}\label{eq:GtCd0}
  ( A_\mu^\prime)^A{}_B  =  R^A{}_C \, (x) \, \Big(  A_\mu - i \partial_\mu   \Big)^C{}_D     R^{\dagger D}{}_B \, (x) \, ,
\end{eqnarray}
where
\begin{eqnarray}\label{eq:}
R^A{}_B  (x) = \Big( e^{- i \Omega (x) }  \Big)^A{}_B     = \delta^A_B - i  (\Omega (x) )^A{}_B  +  \cdots \, ,
\end{eqnarray}

This consideration necessarily introduces a new field: the general connection $(A_\mu (x))^A{}_B $, which appears within the covariant derivative. This connection describes the rule for parallel transport and reflects the geometry of the group manifold. Therefore, it is universal for all fields transforming under that group. This new field is the interacting gauge field and must itself satisfy all the properties derived from the principle field equations.

Finally, we require a coordinate system that is well-defined throughout the entire space. To achieve this, we will retain one index as the space-time coordinate index $\mu$, while preserving all other tangent space indices
$(a, b, \cdots)$ contained in the set $A$.

\subsection{Gauge transformations in  the non-abelian case}

Using equation (\ref{eq:GSbh1Dn}),  we can rewrite the transformation law of the covariant derivative
\begin{eqnarray}\label{eq:GtCd}
  ({\cal D}_\mu^\prime)^A{}_B  =  R^A{}_C \, (x) \, ({\cal D}_\mu)^C{}_D     R^{\dagger D}{}_B \, (x) \,   .
\end{eqnarray}
Then,  the transformation law for the commutator of covariant derivatives follows directly
\begin{eqnarray}\label{eq:}
[{\cal D}^\prime_\mu, {\cal D}^\prime_\nu]^A{}_B = R^A{}_C (x)   [{\cal D}_\mu, {\cal D}_\nu]^C{}_D  R^{\dagger D}{}_B (x) \,   .
\end{eqnarray}

 Using (\ref{eq:SpCf}), this leads to the transformation law for the field strength tensor
\begin{eqnarray}\label{eq:}
( {\cal F}^{\prime}{}_{\mu \nu} (A) )^A{}_B
=   R^A{}_C (x)  ({\cal F}_{\mu \nu} (A) )^C{}_D  R^{\dagger D}{}_B (x)     \,  ,
\end{eqnarray}
whose infinitesimal form is
\begin{eqnarray}\label{eq:dFmnl}
\delta ({\cal F}_{\mu \nu} (A) )^A{}_B =  - i [ \Omega,   ({\cal F}_{\mu \nu} (A) )]^A{}_B      \, .
\end{eqnarray}
The same result can be derived from expressions (\ref{eq:FSGe}) and (\ref{eq:GSbh1Dn}).

\subsection{Bilinear form of  the  action for gauge fields  $A_\mu$}

We now construct gauge-invariant combinations of the fields. The square of the field strength tensor transforms as
\begin{eqnarray}\label{eq:}
(   {\cal F}^{\prime}{}^{\mu \nu} (A)   {\cal F}^{\prime}{}_{\mu \nu} (A) )^A{}_B
=   R^A{}_C (x)  (  {\cal F}^{\mu \nu} (A) {\cal F}_{\mu \nu} (A) )^C{}_D  R^{\dagger D}{}_B (x)     \, ,
\end{eqnarray}
which shows that its trace is gauge-invariant. Therefore, a suitable Lagrangian  is
\begin{eqnarray}\label{eq:Lbf}
  {\cal L}  =   - \frac{1}{4}    Tr  \Big(  {\cal F}^{\mu \nu} (A) {\cal F}_{\mu \nu} (A)   \Big)    \,   .
\end{eqnarray}
This action describes a nontrivial interacting field theory. We will present two prominent examples: Yang-Mills theory, which describes massless particles of helicity $\lambda = 1$, and  general relativity, which corresponds to the case of helicity $\lambda = 2$.


\section{Yang-Mills theory - example with helicity   $\lambda = 1$}
\setcounter{equation}{0}

In Yang-Mills theory, local tangent space coordinates  $x^a$  provide a well-defined coordinate system over the entire space. Consequently, we can simply  take    $x^\mu = x^a$.

In this example, we consider a set of vector fields transforming under the   $SU (N)$   gauge group, with generators   $(t_I)^A{}_B$.    Following equation (\ref{eq:VfnAg}), we can introduce the gauge field
\begin{eqnarray}\label{eq:VfnAg1}
  (A_\pm^a)^A{}_B   =    (V_{\pm 1})^{a I}  (t_I)^A{}_B
  =  \breve{p}_\pm^a  \mathbf{V}_\mp^I  (t_I)^A{}_B    \, ,
\end{eqnarray}
and  the gauge parameter
\begin{eqnarray}\label{eq:OAB}
 (\Omega)^A{}_B = \Omega^I (t_I)^A{}_B    \, .
\end{eqnarray}

The commutator of the fields takes the form
\begin{eqnarray}\label{eq:Com2}
[A_a (x), A_b  (x)]^A{}_B
=   A^I_a   A^J_b    ([t_I, t_J ])^A{}_B  =  A^I_a   A^J_b   f_{I J}{}^K  (t_K )^A{}_B      \, ,
\end{eqnarray}
and the commutator between the field and the gauge parameter is
\begin{eqnarray}\label{eq:Com2o}
[A_a (x), \Omega (x)]^A{}_B
=   A^I_a \Omega^J   ([t_I, t_J ])^A{}_B  =  A^I_a  \Omega^J  f_{I J}{}^K  (t_K )^A{}_B   \, ,
\end{eqnarray}
where  $f_{I J}{}^K $ are the structure constants of the group.

The gauge transformation is therefore given by
\begin{eqnarray}\label{eq:DAmuPf}
\delta (A_a  (x) )^A{}_B    =  ( \partial_a \Omega  +  i [A_a, \Omega] )^A{}_B
=   D_a \Omega^I   (t_I )^A{}_B     \, ,
\end{eqnarray}
where we have introduced the standard covariant derivative
\begin{eqnarray}\label{eq:}
D_a  \Omega^I   = \partial_a \Omega^I   + i f^I{}_{J K} A_a^J \Omega^K    \, .
\end{eqnarray}

The general expression for the field strength
\begin{eqnarray}\label{eq:RabmnY}
( {\cal F}_{a b} (A) )^A{}_B = \Big( \partial_a A_b  - \partial_b  A_a  + i [ A_a ,  A_b ] \Big)^A{}_B     \, ,
\end{eqnarray}
assumes this well-known form for the  $SU (N)$   group
\begin{eqnarray}\label{eq:RabmnY1}
( {\cal F}_{a b } (A) )^A{}_B =   F_{a b}^I  (t_I)^A{}_B \, ,
\end{eqnarray}
where
\begin{eqnarray}\label{eq:}
  F_{a b }^I  =  \partial_a  A_b^I - \partial_b  A_a^I  + i   f^I{}_{J K}   A_a^J  A_b^K      \, .
\end{eqnarray}

Finally, according to equation (\ref{eq:Lbf}), the Yang-Mills Lagrangian is proportional to the trace of the square of the field strength, which is gauge invariant
\begin{eqnarray}\label{eq:}
 {\cal L}_{YM} =   Tr \Big(  {\cal F}^{a b } (A) {\cal F}_{a b} (A)   \Big)       \,  .
\end{eqnarray}


\section{General relativity as a  local Poincare  gauge theory:  the helicity $\lambda = 2$  example}
\setcounter{equation}{0}

The second example is  general relativity formulated as a local Poincare gauge theory.

\subsection{Gauge fields and gauge parameters  for the Poincare  group}

We denote the generators, corresponding   connections, and group parameters of the Poincare group as follows
\begin{eqnarray}\label{eq:}
(t_I)^A{}_B = \{(P_a)^A{}_B, (M_{a b})^A{}_B \} \, ,  \qquad  A^I_\mu = \{B_\mu^a, - \frac{1}{2} \omega_\mu^{a b}  \}   \, ,
   \qquad  \Omega^I = \{\varepsilon^a, - \frac{1}{2} \omega^{a b} \} \, ,
\end{eqnarray}
where    $I= \{ a, [a b] \}$.  Note that  $M_{a b}$,  $ \omega_\mu^{a b}$   and  $\omega^{a b}$
are antisymmetric in the indices    $a$ and $b$.

We can introduce the general Poincare  connection
\begin{eqnarray}\label{eq:Amuc}
(A_\mu)^A{}_B  =  A^I_\mu (t_I)^A{}_B =  B_\mu^a (x ) (P_a)^A{}_B - \frac{1}{2} \omega_\mu^{a b} (x) (M_{a b})^A{}_B   \, ,
\end{eqnarray}
whose components are the spin connection   $\omega_\mu^{a b} (x)$    and the translation gauge potential  $B_\mu^a (x)$.  The general gauge parameter is
\begin{eqnarray}\label{eq:OAB1}
 (\Omega)^A{}_B = \Omega^I (t_I)^A{}_B =  \varepsilon^a (x ) (P_a)^A{}_B - \frac{1}{2} \omega^{a b} (x) (M_{a b})^A{}_B \, .
\end{eqnarray}

Let us explain the connection with the previous   notations.   Similar to the relation    (\ref{eq:GSbh1v})  we have
\begin{eqnarray}\label{eq:GSbh1bo}
 B^a_\mu   (k ) =   \alpha   (B^a_\mu)_{+ 2}  (k )  +  \beta  ( B^a_\mu)_{- 2}  (k )  \,   , \qquad
    \omega_\mu^{a b}   (k ) =   \alpha   (\omega_\mu^{a b} )_{+ 2}  (k )  +  \beta  ( \omega_\mu^{a b} )_{- 2}  (k )  \,
\end{eqnarray}
where
\begin{eqnarray}\label{eq:GSbh1bo1}
   (B^a_\mu)_{\pm 2}  (k )     =   e_{  \mu b}  (T_{\pm 2})^{c b}    (k )  \,   , \qquad
  - \frac{1}{2} (\omega_\mu^{a b} )_{ \pm  2}  (k )   =    e_{ \mu c}    (T_{\pm 2})^{[a b] c}         (k )  \,  .
\end{eqnarray}

These fields form an incomplete irreducible representation of the extended group that includes spatial inversion.    The representations  $  (B^a_\mu)_{+ 2}^a  (k)$  and $ (B^a_\mu)_{-2}^a  (k )$  as well as
$ (\omega_\mu^{a b} )_{+ 2}^a  (k)$  and $ (\omega_\mu^{a b} )_{-2}^a  (k )$    are one-dimensional, meaning each possesses  a single   degree of freedom. The combined fields   $  B^a_\mu  (k)$  and
$\omega_\mu^{a b}   (k ) $  therefore carry  two degrees of freedom and  describe the  graviton.

To construct  the real fields  in  the coordinate representation, we first boost from the standard momentum frame to an arbitrary frame.   We then obtain the  coordinate-dependent fields     $  B^a_\mu  (x)$  and
$\omega_\mu^{a b}   (x ) $  via the mapping    $k^a  \to p^a \to    i \partial^a$,   analogous to the procedure in electrodynamics given by Eq. (\ref{eq:Coore}).

Following the same logic, we introduce the general gauge parameter in the coordinate representation   $ \Omega^A{}_B (x) $.

\subsection{Field strength for the Poincare   group: torsion and  curvature}

Using the general expression for the covariant derivative    (\ref{eq:CodegA}), we can obtain the corresponding form for the Poincare group
\begin{eqnarray}\label{eq:CodegA1}
({\cal D}_\mu)^A{}_B = \delta^A_B \partial_\mu + i  [A_\mu (x) ]^A{}_B
=  \delta^A_B \partial_\mu  + i  B_\mu^a (x ) (P_a)^A{}_B - \frac{i}{2} \omega_\mu^{a b} (x) (M_{a b})^A{}_B   \, .
\end{eqnarray}

The general field strength is given by Eq. (\ref{eq:FSGe}). To calculate it  in case of  Poincare  group, we first derive the commutator   $[A_\mu (x), A_\nu (x)]$   using the connection (\ref{eq:Amuc}) and the Poincare Lie algebra  (\ref{eq:PG})
\begin{eqnarray}\label{eq:Com21}
[A_\mu (x), A_\nu (x)]^A{}_B
=  i \Big[ - \omega_\mu^a{}_b  (x) B_\nu^b (x ) +  \omega_\nu^a{}_b (x)   B_\mu^b (x ) \Big] (P_a)^A{}_B    \nonumber \\
+ \frac{i}{2} \Big[ \omega_\mu^a{}_c (x)   \omega_\nu^{c b} (x) - \omega_\nu^a{}_c (x)   \omega_\mu^{c b} (x) \Big]  (M_{a b})^A{}_B   \, .
\end{eqnarray}

Therefore, from the definition     (\ref{eq:FSGe}), we obtain
\begin{eqnarray}\label{eq:PIA27}
 ({\cal F}_{\mu \nu} (x) )^A{}_B   =  T_{\mu \nu}^a   (P_a)^A{}_B
- \frac{1}{2}  R^{a b}{}_{\mu \nu} (\omega)  (M_{a b})^A{}_B    \, ,
\end{eqnarray}
where we have introduced the torsion
\begin{eqnarray}\label{eq:T}
T_{\mu \nu}^a = D_\mu B_\nu^a   - D_\nu B_\mu^a \, , \qquad   D_\mu B_\nu^a = \partial_\mu B_\nu^a   +\omega_\mu^a{}_b  B_\nu^b    \, ,
\end{eqnarray}
and    curvature
\begin{eqnarray}\label{eq:R}
R^{a b}{}_{\mu \nu} (\omega) = \partial_\mu \omega_\nu^{a b} - \partial_\nu \omega_\mu^{a b} +  \omega_\mu^{a c}  \omega_{\nu c}{}^b -  \omega_\nu^{a c}  \omega_{\mu c}{}^b \, .
\end{eqnarray}

An alternative way to obtain these relations is through direct calculation using the general expression for the field strength    (\ref{eq:SpC})
\begin{eqnarray}\label{eq:SpCfa}
[{\cal D}_\mu,  {\cal D}_\nu]^A{}_B  \Psi^B
= i T_{\mu \nu}^a (P_a)^A{}_B    \Psi^B   - \frac{i}{2} R^{a b}{}_{\mu \nu} (\omega)    (M_{a b})^A{}_B  \Psi^B  \,  .
\end{eqnarray}

\subsection{Local gauge transformation for  connections,  torsion and curvature   in  the Poincare   group}

The gauge transformation of the connection,   $A_\mu (x)$, is governed by the commutator
\begin{eqnarray}\label{eq:Com2o2}
[A_\mu (x), \Omega (x)]^A{}_B
=  i \Big[ - \omega_\mu^a{}_b  (x) \varepsilon^b (x )  +  \omega^a{}_b (x)   B_\mu^b (x ) \Big] (P_a)^A{}_B     \nonumber \\
+ \frac{i}{2} \Big[ \omega_\mu^a{}_c (x)   \omega^{c b} (x) +   \omega_\mu^{b}{}_c (x) \omega^{a c} (x) \Big]  (M_{a b})^A{}_B   \, .
\end{eqnarray}

From this, we can derive the local gauge transformations for the gauge potentials within the Poincare group
\begin{eqnarray}\label{eq:DAmuPf1}
\delta (A_\mu  (x) )^A{}_B    =  ( \partial_\mu \Omega  +  i [A_\mu, \Omega] )^A{}_B     \nonumber \\
=  \Big[  D_\mu \varepsilon^a (x) - \omega^a{}_b   (x)  B_\mu^b  (x)  \Big] (P_a)^A{}_B  - \frac{1}{2}  D_\mu \omega^{a b} (x)  (M_{a b})^A{}_B   \, ,
\end{eqnarray}
 Introducing the  standard covariant derivative
\begin{eqnarray}\label{eq:}
 D_\mu \varepsilon^a (x) =  \partial_\mu \varepsilon^a (x ) +  \omega_\mu^a{}_b  (x) \varepsilon^b (x ) \, , \nonumber \\
D_\mu \omega^{a b} (x) =  \partial_\mu \omega^{a b} (x) + \omega_\mu^a{}_c (x)   \omega^{c b} (x) + \omega_\mu^b{}_c (x) \omega^{a c} (x) \,  ,
\end{eqnarray}
the local gauge transformations for the component fields are found to be
\begin{eqnarray}\label{eq:lLgtp1faf}
\delta_\varepsilon B_\mu^{a} (x)  = D_\mu \varepsilon^a (x)   \, , \qquad
\delta_\omega B_\mu^{a} (x)  =    - \omega^a{}_b   (x)  B_\mu^b  (x) \, ,  \nonumber \\
\delta_\varepsilon  \omega_\mu^{a b} (x)  =  0 \, , \qquad
\delta_\omega  \omega_\mu^{a b} (x) = D_\mu \omega^{a b} (x) \, .
\end{eqnarray}

Applying the general expression for the local gauge transformation of the field strength (Eq. \ref{eq:dFmnl}) to the specific case of the Poincare  group, we obtain
\begin{eqnarray}\label{eq:}
\delta_\omega  T^a{}_{\mu \nu} =  - \omega^a{}_c T^c{}_{\mu \nu}  \,  , \qquad
\delta_\varepsilon  T^a{}_{\mu \nu} =    R^{a b}{}_{\mu \nu} (\omega)  \varepsilon_b  \, , \nonumber \\
\delta_\omega  R^{a b}{}_{\mu \nu} (\omega) =    \omega^a{}_c (x)   R^{c b}{}_{\mu \nu} (\omega) + \omega^b{}_c (x)   R^{a c}{}_{\mu \nu} (\omega)  \, ,  \qquad
\delta_\varepsilon  R^{a b}{}_{\mu \nu} (\omega) =  0 \, .
\end{eqnarray}

Consequently, both the torsion  $  T^a{}_{\mu \nu} $ and the curvature  $ R^{a b}{}_{\mu \nu}$ transform as Lorentz tensors with respect to their   $a$ and $b$  indices.

It is important to emphasize that the expressions for the field strength and its local gauge transformations are formally identical for both Yang-Mills theory (with helicity $\lambda = 1$) and general relativity
(with helicity $\lambda = 2$). The fundamental distinction lies in the structure constants of the underlying gauge group.

\subsection{Tetrad field     }

The   field    $\Psi^A [x^a (x^\mu)] $    is a function of the local, tangent space coordinates $x^a$, which in turn depend on the spacetime coordinates  $x^\mu$. The tangent space is a flat Minkowski space at each point of the manifold.

The key insight is to treat the tangent space coordinates $x^a$  not just as coordinates, but as a vector field $x^a (x^\mu)$   in the   flat space.  Using definition  (\ref{eq:Dcodg})  we can then apply the  covariant derivative to this vector field  $x^a$ instead to  field $\Psi^A$
\begin{eqnarray}\label{eq:}
d x^\mu \equiv \varepsilon n^\mu \, , \qquad
D x^a \equiv x^a  (x^\mu + \varepsilon n^\mu ) - x_\|^a  (x^\mu + \varepsilon n^\mu ) = x^a  (x^\mu + d x^\mu ) - x_\|^a  (x^\mu + d x^\mu ) \,  . \qquad
\end{eqnarray}

Then from  definition  (\ref{eq:Dcodg}) we have
\begin{eqnarray}\label{eq:Dcodga}
D x^a    = d x^\mu  ({\cal D}_\mu x)^a  \, .
\end{eqnarray}
Using expression  for covariant derivatives   for the Poincare group (\ref{eq:CodegA1})    we obtain
\begin{eqnarray}\label{eq:Dcodga1}
D x^a =  d x^\mu  \Big[ \delta^a_b \partial_\mu + i  B_\mu^c (x) (P_c)^a{}_b  - \frac{i}{2} \omega_\mu^{c d} (x) (M_{c d})^a{}_b  \Big] x^b  \, .
\end{eqnarray}

Since,  vector representations of Poincare generators are
\begin{eqnarray}
(P_c)^a{}_b  =  \delta^a_b  i \partial_c  \, ,      \nonumber \\
(M_{c d})^a{}_b = (S_{c d})^a{}_b + \delta^a_b  L_{c d}
=  i \Big( \delta^a_c \, \eta_{d b}  - \delta^a_d \, \eta_{b c}  \Big)  + i \delta^a_b (x_c \partial_d - x_d \partial_c)  \, ,
\end{eqnarray}
we obtain
\begin{eqnarray}\label{eq:}
 (M_{c d})^a{}_b x^b = 0 \, .
\end{eqnarray}
Substituting this back into the covariant derivative of $x^a$ we have
\begin{eqnarray}\label{eq:Dcodga2}
D x^a  =  d x^\mu \Big(  \partial_\mu x^a -  B_\mu^a (x)     \Big)
=  d x^a    -  B_\mu^a (x)  d x^\mu   \, .
\end{eqnarray}

Defining tetrads  $e^a{}_\mu$   with relation
\begin{eqnarray}\label{eq:Dcodga3}
 D x^a  = e^a{}_\mu d x^\mu   \, ,
\end{eqnarray}
 we obtain
\begin{eqnarray}\label{eq:Dcodga4}
 e^a{}_\mu  =   \partial_\mu x^a -  B_\mu^a (x)      \, .
\end{eqnarray}
It connect   local, tangent space  coordinates $x^a$ with  space-time  coordinates $x^\mu$ in the case of local Poincare theory.

The metric tensor takes the  form
\begin{eqnarray}\label{eq:Dcodga5}
   g_{\mu \nu } =  \eta _{a b}  e^a{}_\mu   e^b{}_\nu
=    \eta _{a b}  \Big( \partial_\mu x^a -  B_\mu^a (x) \Big)   \Big( \partial_\nu x^b -  B_\nu^b (x) \Big)  \, .
\end{eqnarray}
It confirms that the tetrad is indeed the fundamental field that describes the geometry, as the metric is a derived quantity.


\section{Gauge invariant Lagrangians for general relativity}
\setcounter{equation}{0}

We have previously constructed a gauge-invariant expression bilinear in the field strength (see Eq. \ref{eq:Lbf}). For the local Poincare group, this takes the form
\begin{eqnarray}\label{eq:}
{\cal L}_2  = Tr \Big( R^{a b}{}_{\mu \nu} S_{a b} \Big)^2  =  4  R^{\mu \nu \rho \sigma}
R_{\mu \nu \rho \sigma}   \,   .
\end{eqnarray}

\subsection{Palatini formalism}

For the local Poincare group, an alternative method exists for constructing gauge invariants linear in the field strength, which leads to the scalar curvature. The general form of such a gauge invariant expression is
\begin{eqnarray}\label{eq:PIA27vl}
   {\cal L} =  i ({\cal F}_{\mu \nu} (x) )^c{}_d   e^\mu{}_c \, e^{\nu d}   = i T_{\mu \nu}^a   (P_a)^c{}_d   e^\mu{}_c \, e^{\nu d}
- \frac{i}{2}  R^{a b}{}_{\mu \nu} (\omega)  (S_{a b})^c{}_d  \,  e^\mu{}_c \, e^{\nu d}    \,  .
\end{eqnarray}
Using the expressions for the translation and spin generators in the vector representation,
\begin{eqnarray}\label{eq:SOvfg1}
   (P_a)^c{}_d = \delta^c_d \partial_a \, ,   \qquad
(S_{a b})^c{}_d  =  i \Big(    \delta^c_a   \eta_{b d} - \delta^c_b   \eta_{a d}  \Big)     \,   ,
\end{eqnarray}
we find
\begin{eqnarray}\label{eq:PIA27vl1}
  {\cal L} =   i  T_{\mu \nu}^a  \partial_a  g^{\mu  \nu}    +  R   (\omega)             \,  ,
\end{eqnarray}
where $R (\omega)$  is a scalar curvature.
The  first term   is zero  because $ T_{\mu \nu}^\rho  $ is antisymmetric  in  $\mu$  and  $\nu$, while
 $ g^{\mu  \nu} $    is symmetric under   their exchange.

Consequently, the linear gauge-invariant Lagrangian is independent of torsion, and we obtain
\begin{eqnarray}\label{eq:PIA27vl2}
  {\cal L} =       R   (\omega)   \,  .
\end{eqnarray}
The corresponding    action,
\begin{eqnarray}\label{eq:}
S_P (e,  \omega)  =   \int  d^4 x e R  ((\omega) =   \int  d^4 x e      R^{a b}{}_{\mu \nu} (\omega)       e^\mu{}_a \, e^\nu{}_b
\,    \qquad   e  = \det  e^a{}_\mu \, ,
\end{eqnarray}
 is  the   Palatini formulation   of general relativity,   \cite{P}.

Using the identity
\begin{eqnarray}\label{eq:}
  \varepsilon_{a b c d}   e^a_\mu e^b_\mu  e^c_\rho  e^d_\sigma
= e  \varepsilon_{\mu \nu \rho \sigma}   \, ,
\end{eqnarray}
we   find
\begin{eqnarray}\label{eq:}
  \varepsilon_{a b c d}   e^a_\mu e^b_\mu
= e  \varepsilon_{\mu \nu \rho \sigma}    e_c^\rho    e_d^\sigma          \,  .
\end{eqnarray}
 Multiplying  this  by the Levi-Civita tensor $\varepsilon^{\mu \nu \rho_1 \sigma_1}$   yields
\begin{eqnarray}\label{eq:}
e    \,   e^\mu_a \, e^\nu_b      = -  \frac{1}{4}    e^c_\rho  e^d_\sigma
 \varepsilon_{a b c d} \varepsilon^{\mu \nu \rho \sigma}   \,  .
\end{eqnarray}

Therefore, the Palatini action can be rewritten as
\begin{eqnarray}\label{eq:}
S_P (e,  \omega)  =   -  \frac{1}{4}   \varepsilon_{a b c d} \varepsilon^{\mu \nu \rho \sigma}
   \int  d^4 x        R^{a b}{}_{\mu \nu} (\omega)       e^c_\rho \, e^d_\sigma      \,  ,
\end{eqnarray}
where the curvature tensor is given by
\begin{eqnarray}\label{eq:Ro5}
R^{a b}{}_{\mu \nu} (\omega) = \partial_\mu \omega_\nu^{a b} - \partial_\nu \omega_\mu^{a b}  + [\omega_\mu , \omega_\nu]^{a b}  \,  .
\end{eqnarray}

\subsection{ Self-dual   spin connection }

To prepare the notation for coupling to chiral spinors and to introduce the Ashtekar formalism, we will define the self-dual and anti  self-dual spin connections. We begin by introducing the dual spin connection
\begin{eqnarray}\label{eq:}
{}^\star \omega_\mu^{a b}  \equiv \frac{1}{2} \, \varepsilon^{a b}{}_{c d} \, \omega_\mu^{c d} \, .
\end{eqnarray}
From this, we can form the linear combinations
\begin{eqnarray}\label{eq:odsd1}
\omega_{\pm \mu}^{a b} = \frac{1}{2} ( \omega_\mu^{a b} \mp i \, {}^\star \omega_\mu^{a b} ) \,  .
\end{eqnarray}

For a Minkowski signature, the dual   spin connection  satisfies  relation ${}^{\star \star } \omega_\mu^{a b} = - \omega_\mu^{a b}$   which gives
\begin{eqnarray}\label{eq:}
{}^\star \omega_{\pm \mu}^{a b} = \pm \, i \, \omega_{\pm \mu}^{a b} \, .
\end{eqnarray}
Consequently,  $\omega_{ + \mu}^{a b}$ is self-dual and $\omega_{ - \mu}^{a b}$ is anti self-dual.

Based on Eq.eq.(\ref{eq:odsd1}), we can define projection operators onto the self-dual and anti self-dual parts of the spin connection
\begin{eqnarray}\label{eq:}
\Pi_\pm = \frac{1}{2} (1 \mp i \, {}^\star) \,  .
\end{eqnarray}
Using the property   $({}^\star)^2 = -1$ we find  that $\Pi_\pm$  are indeed  projectors
\begin{eqnarray}\label{eq:}
  \Pi_\pm^2 = \Pi_\pm \,  , \qquad \Pi_+ \Pi_- =0 \,  , \qquad \Pi_+ + \Pi_-  = 1 \, .
\end{eqnarray}

Expressed explicitly with indices, we have
\begin{eqnarray}\label{eq:CHpr}
\omega_\pm^{a b} =  \Pi_\pm {}^{a b}{}_{c d}  \, \omega^{c d} \,  , \qquad
\Pi_\pm {}^{a b}{}_{c d} = \frac{1}{4} (\delta^a_c \, \delta^b_d  - \delta^a_d \, \delta^b_c \mp i \, \varepsilon^{a b}{}_{c d} )  = \Pi_\pm {}_{c d} {}^{a b} \,  , \nonumber \\
1 \sim  \frac{1}{2} \, (\delta^a_c \, \delta^b_d  - \delta^a_d \, \delta^b_c)   \, , \qquad    {}^\star  \sim  \frac{1}{2} \, \varepsilon^{a b}{}_{c d}   \, .
\end{eqnarray}

\subsection{ Ashtekar formalism in terms of  self-dual curvature}

Using the methods developed in this article, we will derive the Ashtekar action \cite{A1, A2}.

First, we establish a useful identity.
Let $F^{a b}$ and $G^{a b}$  be tensors with two Lorentz indices. We defined Lie brackets as
\begin{eqnarray}\label{eq:}
[F , G]^{a b} \equiv F^a{}_c G^{c b} - G^a{}_c F^{c b}  \,  .
\end{eqnarray}
It produces
\begin{eqnarray}\label{eq:}
({}^\star [F , {}^\star G ])^{a b} = - [F , G]^{a b} \,  ,
\end{eqnarray}
and consequently, for the projections
\begin{eqnarray}\label{eq:}
\Pi_ \pm [F , G] = [F , G_\pm] = [F_\pm , G] = [F_\pm , G_\pm] \,  , \qquad  [F_\pm , G_\mp] = 0  \, ,
\end{eqnarray}
where $F_\pm \equiv \Pi_\pm F$. It follows that
\begin{eqnarray}\label{eq:fgs}
[F , G] = [F_+ , G_+] + [F_- , G_-]\, .
\end{eqnarray}

Using this relation, the curvature of the spin connection  $\omega_\mu^{a b}$
\begin{eqnarray}\label{eq:Ro1}
R^{a b}{}_{\mu \nu} (\omega) = \partial_\mu \omega_\nu^{a b} - \partial_\nu \omega_\mu^{a b}  + [\omega_\mu , \omega_\nu]^{a b}  \,  ,
\end{eqnarray}
can be   decomposed  into self-dual and anti self-dual parts
\begin{eqnarray}\label{eq:}
R^{a b}{}_{\mu \nu} (\omega) = R^{a b}{}_{\mu \nu} (\omega_+) + R^{a b}{}_{\mu \nu} (\omega_-) \, .
\end{eqnarray}

These components are given explicitly by the projections
\begin{eqnarray}\label{eq:Ropm}
R^{a b}{}_{\mu \nu} (\omega_\pm) = \Pi_\pm R^{a b}{}_{\mu \nu} (\omega) = \frac{1}{2} \left[R^{a b}{}_{\mu \nu} (\omega)   \mp i {}^\star  R^{a b}{}_{\mu \nu} (\omega) \right] \,  .
\end{eqnarray}

We   now define the vector field $A_\mu^{a b}$ as the self-dual spin connection
\begin{eqnarray}\label{eq:}
A_\mu^{a b} \equiv \omega_{+ \mu}^{a b}
=     \frac{1}{2} ( \omega_\mu^{a b}  -  i \, {}^\star \omega_\mu^{a b} )      \,  ,
\end{eqnarray}
 which, by construction, satisfies
\begin{eqnarray}\label{eq:}
{}^\star    A_\mu^{a b} =  i \,  A_\mu^{a b} \, .
\end{eqnarray}

The corresponding field strength is the self-dual part of the curvature
\begin{eqnarray}\label{eq:FS}
F^{a b}{}_{\mu \nu} (A) = R^{a b}{}_{\mu \nu} (\omega_+) = \frac{1}{2} \, \left[R^{a b}{}_{\mu \nu} (\omega)   - i \, {}^\star  R^{a b}{}_{\mu \nu} (\omega) \right] \,  ,
\end{eqnarray}
or, written explicitly,
\begin{eqnarray}\label{eq:}
F^{a b}{}_{\mu \nu} (A) = \frac{1}{2} \, \left[R^{a b}{}_{\mu \nu} (\omega)   - \frac{i}{2}\, \varepsilon^{a b}{}_{c d}  \, R^{c d}{}_{\mu \nu} (\omega) \right] \,  .
\end{eqnarray}

We now introduce the Ashtekar action
\begin{eqnarray}\label{eq:}
S_{As}  (e, A)  =    \int  d^4 x e      F^{a b}{}_{\mu \nu} ( A)       e^\mu{}_a \, e^\nu_b
=    \varepsilon^{\mu \nu \rho  \sigma }   \int  d^4 x       F^{a b}{}_{\mu \nu} ( A)
e_{\rho a} \, e_{ \sigma b}                              \, .
\end{eqnarray}

This action can be expressed in terms of the Palatini action as
\begin{eqnarray}\label{eq:}
S_{As}  (e, A)  =    S_P (e, \omega)  - i   T (e, \omega)               \, ,
\end{eqnarray}
where
\begin{eqnarray}\label{eq:}
T  (e,  \omega )  =    \int  d^4 x e   \,  {}^\star  R^{a b}{}_{\mu \nu} (\omega)       e^\mu{}_a \, e^\nu_b    \, .
\end{eqnarray}

Since $T  (e,  \omega )$  is a topological  term   that  does not affect the classical  equations of motion,    Ashtekar action is equivalent to  the Palatini action  and,  consequently, to the Hilbert-Einstein action.

Thus, in the Ashtekar formulation, the fundamental dynamical variable is the self-dual connection $A_\mu^{a b}$.


\section{Dirac equation in curved space-time}
\setcounter{equation}{0}

Since matter fields are predominantly fermionic, we now formulate the Dirac equation in curved spacetime. A primary challenge is that standard partial derivatives are not well-defined on a curved manifold. To address this, we employ the Fock-Ivanenko covariant derivative, the appropriate differential operator, which is defined in Appendix \ref{sec:FID}.

Following \cite{FI, F}, we define the operator $F$ as
\begin{eqnarray}\label{eq:}
F \psi \equiv (i \gamma^\mu    {\cal D_\mu}   - m) \psi    \, ,     \qquad
\end{eqnarray}
where the Fock-Ivanenko covariant derivative in the spinor representation is
\begin{eqnarray}\label{eq:}
  ({\cal D_\mu})^\alpha{}_\beta  =   \delta^\alpha_\beta   \partial_\mu + i  (\Omega_\mu )^\alpha{}_\beta    \,  .
\end{eqnarray}

Here, the connection $(\Omega_\mu )^\alpha{}_\beta$, given by Eq. (\ref{eq:Resog}), is
\begin{eqnarray}\label{eq:Reso1d}
( \Omega_\mu )^\alpha{}_\beta = \frac{i}{4} \,  \omega_\mu^{a b} \, ( \sigma_{a b}  )^\alpha{}_\beta  - i  B_\mu^a  (P_a   )^\alpha{}_\beta   - i a_\mu^{I} (\tau_I   )^\alpha{}_\beta  \,  .
\end{eqnarray}

From the Hermitian properties of gamma matrices    and  relation   $ \gamma^0 \Omega^\dagger_\mu \gamma^0 = \Omega_\mu $,   the adjoint of $F \psi$ is
\begin{eqnarray}\label{eq:}
\overline{(F \psi)} = (F \psi)^\dag \gamma^0 = -i \partial_\mu {\bar \psi}  \gamma^\mu
-   {\bar \psi} \,  \Omega_\mu \gamma^\mu   - m {\bar \psi} \,  .
\end{eqnarray}
The difference between the two  expressions  is then
\begin{eqnarray}\label{eq:Relp1}
{\bar \psi} F \psi  -  \overline{(F \psi)} \,  \psi =   i \partial_\mu ({\bar \psi}  \gamma^\mu\psi) - i {\bar \psi}  (\partial_\mu  \gamma^\mu) \psi - {\bar \psi} [\gamma^\mu,  \Omega_\mu ]  \psi\, .
\end{eqnarray}

Employing the relation   $[\gamma^\mu ,  \Omega_\nu] = - i  \mathrel{{\mathop{\mathop{ D_\nu}_{\mathrm{\Gamma}}}}} \gamma^\mu$,     where
$ \mathrel{{\mathop{\mathop{ D_\mu}_{\mathrm{\Gamma}}}}}$    is the standard covariant derivative involving the Christoffel symbols
\begin{eqnarray}\label{eq:CDGo}
\mathrel{{\mathop{\mathop{\, D_\mu}_{\mathrm{\Gamma}}}}}   V^\nu   = \partial_\mu  V^\nu   +  \Gamma^\nu_{\rho \mu}    V^\rho \,  , \qquad
\end{eqnarray}
we simplify Eq. (\ref{eq:Relp1}) to
\begin{eqnarray}\label{eq:Relp0}
{\bar \psi} F \psi  -  \overline{(F \psi)} \,  \psi =  i \mathrel{{\mathop{\mathop{ D_\mu}_{\mathrm{\Gamma}}}}}       ( {\bar \psi}  \gamma^\mu\psi )   \,  .
\end{eqnarray}
This can be written equivalently as
\begin{eqnarray}\label{eq:Relp}
{\bar \psi} F \psi  -  \overline{(F \psi)} \,  \psi = \frac{i} {\sqrt{-g}} \partial_\mu ({\sqrt{-g}} {\bar \psi} \gamma^\mu \psi) \,  .
\end{eqnarray}

Integrating this expression over spacetime yields
\begin{eqnarray}\label{eq:Relp2}
\int d x  {\sqrt{-g}}  \,  \Big( {\bar \psi} F \psi  -  \overline{(F \psi)} \,  \psi  \Big) = i \int d x \, \partial_\mu ({\sqrt{-g}} \, {\bar \psi} \gamma^\mu \psi) =0 \,  .
\end{eqnarray}
This result demonstrates that the operator $F$ is Hermitian ($F^\dagger = F$). Consequently, the equations of motion for ${\bar \psi}$ and $\psi$ are equivalent.

We may therefore take $F \psi = 0$, or explicitly,
\begin{eqnarray}\label{eq:DeI}
(i \gamma^\mu   {\cal D}  -  m) \, \psi= 0 \,  , \qquad
\end{eqnarray}
as the Dirac equation in general relativity.

Furthermore, from Eq. (\ref{eq:Relp0}), the equation of motion $F \psi = 0$ implies the covariant conservation of the current
\begin{eqnarray}\label{eq:}
\mathrel{{\mathop{\mathop{\, D_\mu}_{\mathrm{\Gamma}}}}} j^\mu = 0    \,  ,  \qquad   (j^\mu \equiv {\bar \psi} \gamma^\mu \psi)   \, .
\end{eqnarray}
Note that this current $j^\mu$ is Hermitian, i.e., $j^{\mu\dagger} = j^\mu$.

.
\subsection{Equations  for chiral spinors}

With the help of the relation
\begin{eqnarray}\label{eq:gasi}
\gamma^a \,  \sigma_{b c}  = i \, (\delta^a_b \, \gamma_c - \delta^a_c \, \gamma_b ) -  \varepsilon^a{}_{b c d} \, \gamma^5 \gamma^d  \,  ,
\end{eqnarray}
 we find
\begin{eqnarray}\label{eq:gaOm}
\gamma^\mu \Omega_\mu = -i \gamma^\mu   b_\mu - \frac{1}{2} \, e^\mu_a \, ( \omega_\mu^{a b}   + i \, {}^\star \omega_\mu^{a b}  \gamma^5 ) \, \gamma_b \, .
\end{eqnarray}
 We can derive its useful form by separating the different chiralities
\begin{eqnarray}\label{eq:gmOm}
\gamma^\mu \Omega_\mu = -i \gamma^\mu    b_\mu - \, e^\mu_a \, ( \omega_{ - \mu}^{a b} \, P_+   + \omega_{ + \mu}^{a b} \, P_- )\,  \gamma_b
= \gamma^\mu \left[  - i   b_\mu  -   e^\nu_a  e_{\mu b}\, \left( \omega_{ - \nu}^{a b} \, P_-  + \omega_{ + \nu}^{a b} \, P_+  \right) \right]  \, ,    \qquad  \qquad
\end{eqnarray}
 using the chirality projection operators
\begin{eqnarray}\label{eq:Chpr}
P_\pm = \frac{1}{2} (1 \pm \gamma^5) \,  .
\end{eqnarray}

Introducing the notation    $\omega_{ \pm \mu} \equiv e^\nu_a \,  e_{\mu b} \, \omega_{ \pm \nu}^{a b}$ or $\omega_{\pm}^b \equiv e^\nu_a \, \omega_{ \pm \nu}^{a b} \equiv
\omega_{ \pm a}^{a b}$,   we express    (\ref{eq:gmOm}) in the form
\begin{eqnarray}\label{eq:gmOm1}
\gamma^\mu \Omega_\mu = - \gamma^\mu \left(  i   b_\mu  +   \omega_{ - \mu} \, P_-  + \omega_{ + \mu} \, P_+   \right)  \, .
\end{eqnarray}
Consequently,  Dirac equation  in curve space-time    becomes
\begin{eqnarray}\label{eq:DeII}
\left\{ \gamma^\mu \left[ i \partial_\mu -   b_\mu  + i   e^\nu_a  e_{\mu b}\, \left( \omega_{ - \nu}^{a b} \, P_-  + \omega_{ + \nu}^{a b} \, P_+  \right) \right]  - m \right\} \, \psi = 0   \,   .
\end{eqnarray}

The expression in brackets can be reformulated using the projections on the self-dual and anti-self-dual parts of the spin connection,  $ \Pi_\pm$, defined in Eq. (\ref{eq:CHpr})
\begin{eqnarray}
\omega_{ - \mu}^{a b} \, P_+   + \omega_{ + \mu}^{a b} \, P_- = (\Pi_- P_+ + \Pi_+ P_-)^{a b}{}_{c d} \, \omega_{\mu}^{c d}  \, .
\end{eqnarray}
This combination is itself a projection operator. Indeed, we have the identities
\begin{eqnarray}\label{eq:}
1 = (\Pi_+ + \Pi_-) (P_+ + P_-) = (\Pi_- P_+ + \Pi_+ P_-) + (\Pi_+ P_+ + \Pi_- P_-) \,  ,
\end{eqnarray}
confirming that both bracketed expressions are projectors.

By multiplying Eq. (\ref{eq:DeII}) with   $P_\pm$, we obtain the equations of motion for the chiral spinors $\psi_\pm  =   P_\pm \psi$     in curved spacetime
\begin{eqnarray}\label{eq:DeIIl}
\left[ \gamma^\mu \left(    i \partial_\mu -  b_\mu  + i   e^\nu_a  e_{\mu b}\,  \omega_{ - \nu}^{a b} \,
   \right)  - m \right]  \, \psi_- = 0   \,   ,
\end{eqnarray}
and
\begin{eqnarray}\label{eq:DeIIr}
\left[   \gamma^\mu \left( i \partial_\mu -  b_\mu  + i   e^\nu_a  e_{\mu b}\,  \omega_{ + \nu}^{a b} \,    \right)  - m \right]  \, \psi_+ = 0   \,   .
\end{eqnarray}

Thus, the left-handed and right-handed chiral spinors couple to the anti selfdual and selfdual connections, respectively. Since chiral spinors are the fundamental building blocks of the  standard model, it is natural to adopt the self-dual connection as the primary variable in the theory.

The second equation is particularly useful for describing matter fields in the Ashtekar formalism \cite{A1, A2}, where the connection is selfdual,   $A_{ \mu}^{a b}  =  \omega_{ + \mu}^{a b}$? . In this context, the equation defines the parallel transport of chiral spinors.

Finally, from these equations of motion, we can derive the corresponding actions. The Dirac action takes the form
\begin{eqnarray}\label{eq:DeA1}
S_D =  \int d^4 x \, e \, {\bar \psi}  \, \left\{ \gamma^\mu \left[ i \partial_\mu -  b_\mu  + i   e^\nu_a  e_{\mu b}\, \left( \omega_{ - \nu}^{a b} \, P_-  + \omega_{ + \nu}^{a b} \, P_+  \right) \right]  - m \right\} \, \psi = 0 \,    ,   \qquad
\end{eqnarray}
and the action for a single chiral spinor is
\begin{eqnarray}\label{eq:DeAc}
S_{As} =  \int d^4 x \, e  \, {\bar \psi}  \, \left[   \gamma^\mu
\left(  i \partial_\mu -   b_\mu  + i   e^\nu_a  e_{\mu b}\,  \omega_{ + \nu}^{a b} \, P_+   \right)  - m \right]    \, \psi = 0 \,    .    \qquad
\end{eqnarray}

\section{ Conclusion   and discussion}
\setcounter{equation}{0}

Reformulating a known theory in a new way can be highly productive. For instance, recasting Maxwell's theory into a more convenient framework   reveals new features  and leads directly to the special theory of relativity, with its inherent Poincare  invariance, and to the concept of Abelian local gauge invariance. This process can then be reversed: we can derive Maxwell's theory from these new, more fundamental principles. Furthermore, these principles can be generalized. By replacing the Abelian gauge symmetry with a non-Abelian one, we arrive at a new theory in this case, Yang-Mills theory. Similarly, by requiring invariance under the local Poincare group, we obtain the  general relativity.

This work follows precisely this logic. We start not from Maxwell's theory, but from the two well-established theories of Yang-Mills and general relativity. Our goal is to show that both can be derived from two fundamental assumptions.

The first assumption is the principle  field equations for the Poincare group [1-3], which generates the equations of motion for free fields of arbitrary spin and helicity. Moreover, the requirement of Lorentz invariance implies Abelian local gauge transformations for massless fields [4, 2].

The second assumption concerns the proper definition of derivatives. Both the equations of motion for free fields and the local gauge transformations involve derivatives. In non-trivial cases where derivatives act on tensors and spinors, we must use Fock-Ivanenko covariant derivatives [5, 6]. These necessarily introduce a new field, the connection, which physically describes interaction. Consequently, from these two principles, we can derive the complete forms of all known equations, including Yang-Mills theory and general relativity in both the Palatini [7] and Ashtekar formalisms [8, 9].

In general, the connection    $A_\mu$  is a vector field (with spacetime index $\mu$) that transforms under a specific group (e.g., $SU(N)$  for Yang-Mills, the Poincare group for gravity). It is therefore useful to express the connection in terms of the group's generators:  $t_I$  for Yang-Mills, and  $P_a$   and $M_{a b} $
 for general relativity. The coefficients in this expansion are the fields that describe the corresponding theory. Thus, Yang-Mills theory is described by the field   $A_\mu^I$, while gravity is described by two fields:
$B_\mu^a$   and  $\omega_\mu^{a b}$.

We derive the general expression for the connection in two independent ways: from the consistency of parallel transport (Eq. \ref{eq:Resog}) and as a connection for helicities $\lambda=1$ and $\lambda = 2$   (Eq. \ref{eq:MGclg})
\begin{eqnarray}\label{eq:Reso1}
( A_\mu  )^A{}_B  = \frac{i}{4} \,  \omega_\mu^{a b} \, (   M_{a b}  )^A{}_B  - i  B_\mu^a  (P_a)^A{}_B   - i a_\mu^{I} (\tau_I )^A{}_B  \,  .
\end{eqnarray}

The paper presents an original way of deriving these well-known theories.  However,  this is not  the primary contribution of this article, but rather the verification of our basic assumptions. Since Yang-Mills theory and general relativity are long-established, they serve as a   strong  test, confirming that our foundational principles are correct.

The natural next step is to generalize these assumptions to obtain new results. For example, instead of the Poincare group, we could begin with a more general group containing it. Starting from the conformal group would initially yield only massless fields.  The subsequent breaking of conformal symmetry down to the Poincare  one  would then introduce masses. This approach provides a potential pathway to understanding the mass spectrum of particles. Within the Poincare group alone, mass is an invariant but arbitrary parameter.  A deeper origin for mass may lie in a broader symmetry.

From the principle field equations for the Poincare group, we obtained equations for both massive and massless free fields. This includes equations for massive bosons and fermions, as well as for massless bosons and their associated local gauge transformations. However, one case remains: that is   massless fermionic fields which was analyzed in  Ref.\cite{S3}.   
The irreducible representations for massless Dirac spinors have been derived, which are represented by states with well defined helicity or chirality. Interactions have been introduced via the appropriate covariant derivatives. Surprisingly, helicity fermions exhibit unconventional interaction properties, because their gravitational interaction does not couple to certain components of the connection. As a result, these particles follow logarithmic spiral trajectories with constant angular velocity, offering a possible explanation for the shape of galaxy rotation curves. For more details, see Ref.\cite{S3}.

\appendix


\section{Parallel transport }
\setcounter{equation}{0}

To compare tensors and spinors at neighboring points    $x^\mu(\tau)$ and  $x^\mu (\tau+ \Delta \tau)$, we must first transport the fields from   $x^\mu(\tau)$   to   $x^\mu (\tau+ \Delta \tau)$. The difference can then be calculated at the same point,   $x^\mu (\tau+ \Delta \tau)$.  This process requires an additional structure known as a connection, which defines how tensors and spinors are transported along a curve.

\subsection{Parallel transport of  vectors }

A vector  $V^\mu$   can be expressed in terms of its projections onto tetrad fields   $e^a_\mu$  as
\begin{eqnarray}\label{eq:Vec}
V^\mu = e^\mu{}_a v ^a \, , \qquad  v ^a = e^a{}_\mu V^\mu  \,  .
\end{eqnarray}

Parallel transport of the vector  is defined  by
\begin{eqnarray}\label{eq:Partr}
V^\mu_\| = V^\mu + \delta V^\mu   \, , \qquad  v^a_\|   =  v^a  + \delta v^a  \,  ,
\end{eqnarray}
 where the infinitesimal changes are given by
\begin{eqnarray}\label{eq:Conn}
\delta V^\mu = - \Gamma^\mu_{\nu\rho} V^\nu d x^\rho  \,  ,  \qquad
\delta v^a = - \omega_\mu {}^a{}_b \,  v^b d x^\mu  \,  .
\end{eqnarray}
Here  $\Gamma^\mu_{\nu\rho}$ is  the  vector connection   and $\omega_\mu {}^a{}_b$ is the Lorentz (spin) connection.

\subsection{Parallel transport of  spinors}

Since we can express tensors as bilinear combinations of the spinors,  parallel transport of the spinors  is connected to   parallel transport of the tensors.   To  find their connection
we are going to  follow    the approach of Ref.\cite{FI, F}.

The parallel transport of a Dirac spinor $\psi$  is defined as
\begin{eqnarray}\label{eq:Delp}
\psi_\| = \psi + \delta \psi  \,  ,  \qquad  \delta \psi = \Omega_\mu \, \psi \, d x^\mu \,  ,
\end{eqnarray}
where   $\Omega_\mu$ is  the spinor connection.
This implies the transport for the  Dirac conjugate spinor
\begin{eqnarray}\label{eq:Delpb}
\delta {\bar \psi} = {\bar \psi} \, \gamma^0  \Omega_\mu^\dag \gamma^0   \, d x^\mu \,  .
\end{eqnarray}

Now, consider a set of  spinor  bilinears
\begin{eqnarray}\label{eq:vA}
t^\Upsilon = {\bar \psi}  \Gamma^\Upsilon  \psi \,  ,
\end{eqnarray}
where  $\Gamma^\Upsilon $    is a set of linearly independent matrices
\begin{eqnarray}\label{eq:basis1}
\Gamma^\Upsilon = \{1, \, \gamma^a, \sigma_{a b} \, , \gamma^5 \gamma^a \,  ,  \gamma^5  \} ,
\end{eqnarray}
 and
\begin{eqnarray}\label{eq:}
    \gamma^5  = i  \gamma^0  \gamma^1  \gamma^2  \gamma^3         \qquad      \sigma_{a b} = \frac{i}{2} \, [\gamma_a , \gamma_b]   \,  .
\end{eqnarray}

From the  Hermiticity   properties of the gamma matrices $\gamma_a^\dag = \gamma^0 \gamma_a \gamma^0$  it follows that
\begin{eqnarray}\label{eq:GH}
\Gamma_\Upsilon^\dag = \gamma^0 \Gamma_\Upsilon \gamma^0  \,  , \qquad  for \qquad
 \Gamma_\Upsilon \neq \gamma^5  \,  ,
\end{eqnarray}
 while for   $\gamma_5$ we have
\begin{eqnarray}\label{eq:g5H}
\gamma_5^\dag =  \gamma_5  =- \gamma^0 \gamma_5 \gamma^0  \,  .
\end{eqnarray}

The parallel transport of spinors must be consistent with the parallel transport of tensors. This means we must find a spinor connection,   $\Omega_\mu$, such that the parallel transport of the  spinor  bilinears
(Eq. \ref{eq:vA}) yields the standard result for tensor parallel transport.

Using Eqs.(\ref{eq:Delp}) and (\ref{eq:Delpb}), we obtain
\begin{eqnarray}\label{eq:dva}
\delta t^\Upsilon = {\bar \psi} \, ( \Gamma^\Upsilon \, \Omega_\mu +  \gamma^0   \Omega_\mu^\dag \gamma^0  \Gamma^\Upsilon) \, \psi  \, d x^\mu \,  .
\end{eqnarray}
 This condition must hold for every index    $\Upsilon$.

Scalars and pseudoscalars remain unchanged under parallel transport. Therefore, for  $\Gamma^\Upsilon  \to 1$, the condition     $\delta ({\bar \psi} \psi) = 0$   implies
\begin{eqnarray}\label{eq:OmH1}
\Omega_\mu + \gamma^0  \Omega_\mu^\dag \gamma^0 = 0 \,  .
\end{eqnarray}
Using this, we can rewrite Eq. (\ref{eq:dva}) as
\begin{eqnarray}\label{eq:dva1}
\delta t^\Upsilon = {\bar \psi} \, [ \Gamma^\Upsilon \, ,\Omega_\mu]  \, \psi  \, d x^\mu \,  .
\end{eqnarray}

For  pseudoscalas, where   $\Gamma^\Upsilon \to \gamma^5$   the condition   $\delta ({\bar \psi} \gamma^5\psi) = 0$   leads to
\begin{eqnarray}\label{eq:}
\gamma^5 \Omega_\mu = \Omega_\mu \gamma^5  \, .
\end{eqnarray}
 We can now expand  $\Omega_\mu$ in the basis Eq.(\ref{eq:basis1}) as
\begin{eqnarray}\label{eq:OM1}
\Omega_\mu =    b_\mu  -   \Omega_\mu^{a b} \sigma_{a b} + i {\bar \Omega}_\mu \gamma^5 \,  .
\end{eqnarray}
Because   $\Omega_\mu$ commutes  with $\gamma^5$,   terms  proportional to  $\gamma^a$ and $\gamma^5 \gamma^a$    are excluded.
Note that  in  accordance with    Eqs.(\ref{eq:GH}), (\ref{eq:g5H}) and (\ref{eq:OmH1}), the coefficients
$b_\mu\, , \Omega_\mu^{a b}$ and ${\bar \Omega}_\mu$    are chosen to be Hermitian.

 For vector fields,  where  $\Gamma^\Upsilon \to  \gamma^a$,  the second relation in Eq.(\ref{eq:Conn}) requires    $\delta ({\bar \psi} \gamma^a\psi) = - \omega_\mu {}^a{}_b \, {\bar \psi} \gamma^b\psi \, d x^\mu $. This imposes the condition
\begin{eqnarray}\label{eq:gaa1}
[\gamma^a ,\Omega_\mu] = -  \omega_\mu {}^a{}_b \gamma^b \, .
\end{eqnarray}
 Substituting the expansion from Eq. (\ref{eq:OM1}) into Eq. (\ref{eq:gaa1}) yields
\begin{eqnarray}\label{eq:}
\Omega_\mu^{a b} =   \frac{1}{4} \,  \omega_\mu^{a b}  \, , \qquad {\bar \Omega}_\mu = 0 \,  .
\end{eqnarray}

Consequently, the final solution is
\begin{eqnarray}\label{eq:Resof}
\Omega_\mu =  - \frac{1}{4} \,  \omega_\mu^{a b} \, \sigma_{a b} +  b_\mu \,  ,
\end{eqnarray}
where $b_\mu$ is solution of the  homogeneous    part of Eq.(\ref{eq:gaa1}).
One can verify that this solution also satisfies the corresponding conditions for
$\Gamma^\Upsilon = \sigma_{a b}$  and   $\Gamma^\Upsilon = \gamma^5 \gamma^a$.

 Note that in addition to the term involving the spin connection  $\omega_\mu^{a b}$, we also have a term with an arbitrary vector field    $b_\mu$.
The only constraint on the field  $b_\mu $   is that it must commute with all matrices $\Gamma^\Upsilon$. We can therefore expand it in terms of Hermitian generators that commute with all $\Gamma^\Upsilon$, such as the generators  $(\tau^I)^\alpha{}_\beta$  of an internal symmetry group (e.g., $SU (N)$   or the translation generators of the Poincare  group   $(P_a)^\alpha{}_\beta$  (which are simply derivatives) in spior representation
\begin{eqnarray}\label{eq:Partrs}
  (b_\mu  )^\alpha{}_\beta  =  a_\mu^{I} (\tau_I )^\alpha{}_\beta   +   B_\mu^a  (P_a)^\alpha{}_\beta   \,  .
\end{eqnarray}

Thus, the most general solution for the connection is
\begin{eqnarray}\label{eq:Resog}
( \Omega_\mu  )^\alpha{}_\beta  = - \frac{1}{4} \,  \omega_\mu^{a b} \, (   \sigma_{a b}  )^\alpha{}_\beta
 +  B_\mu^a  (P_a)^\alpha{}_\beta   +  a_\mu^{I} (\tau_I )^\alpha{}_\beta  \,  .
\end{eqnarray}
 That is a  particular  case of  Eq.(\ref{eq:Amuc})  in spior representation,  where  the spin generator for fermions  is $ S_{a b}  = \frac{1}{2}   \sigma_{a b}$.

The field $ \omega_\mu^{a b}$  is the spin connection, and $ B_\mu^a$ is  the non-trivial part of the tetrad.  Together, they describe the gravitational interaction.
 The field $a^{a I}$  for a suitable choice of the group describes  the electroweak   interaction
(for $U (1)  \times SU (2)$)   and the strong interaction  (for  $SU (3)$).

It is remarkable that the requirement of consistency between the parallel transport of spinors and tensors uniquely determines the form of the gauge fields for all fundamental interactions in nature.

It is important to emphasize   that the parallel transport of spinors is inherently nontrivial. Even when the connection for vectors vanishes   $ \omega_\mu^{a b} = 0$, a nontrivial contribution arises from the homogeneous part of the spinor connection, as shown in (\ref{eq:Partrs}).


\section{Covariant derivatives}\label{sec:FID}
\setcounter{equation}{0}

The covariant derivative is the only consistently definable derivative for non-scalar fields, such as vectors and spinors, on a curved spacetime. We define it using the standard concept of a derivative as the difference between a field's value at  $x^\mu(\tau)$ and  $x^\mu (\tau+ \Delta \tau)$, divided by  $\Delta \tau$. However, since we can only directly add vectors or spinors at the same point in the spacetime manifold, we must first parallel transport one of them to the other's location.

\subsection{Fock - Ivanenko covariant derivatives}

A field    $\Psi^A [x^a (x^\mu)] $   is generally a function of local tangent space coordinates  $x^a$, which themselves depend on the spacetime coordinates    $x^\mu$. To define a derivative, we require a coordinate system that covers the entire spacetime.   We therefore choose to work with the spacetime coordinates $x^\mu$.

To subtract fields defined at different points, we first parallel transport the field   $\Psi^A (x)$  from point
 $x^\mu$ to point  $y^\mu$
\begin{eqnarray}\label{eq:PaTr}
\Psi_\|^A  (y ) =   \Pi^A{}_B  (y, x ) \Psi^B (x) \, ,
\end{eqnarray}
introducing  the comparator $\Pi^A{}_B  (y, x )$.
For an infinitesimal separation    this becomes
\begin{eqnarray}\label{eq:PiABigi}
\Pi^A{}_B  (x^\mu + \varepsilon n^\mu, x^\mu ) = \delta^A_B - i \varepsilon n^\mu (A_\mu (x))^A{}_B  + \cdots  \, ,
\end{eqnarray}
where $(A_\mu)^A{}_B  (x)$ is  the  general connection.

\subsubsection{Fundamental representation  }

We first consider fields in the fundamental representation.
The covariant derivative is defined as the difference between the field  $\Psi^A$ at point $x^\mu + \varepsilon n^\mu $ and parallel transport of the field  $\Psi^A$
from point $x$ to  point $x^\mu + \varepsilon n^\mu $.  Using equations
 (\ref{eq:PaTr})  and       (\ref{eq:PiABigi}) we obtain
\begin{eqnarray}\label{eq:Dcodg}
n^\mu ({\cal D}_\mu \Psi)^A &= &\lim_{\varepsilon \to 0}  \frac{1}{\varepsilon}\, \Big[\Psi^A  (x^\mu + \varepsilon n^\mu ) - \Psi_\|^A  (x^\mu + \varepsilon n^\mu ) \Big]  \nonumber \\
&=& \lim_{\varepsilon \to 0}  \frac{1}{\varepsilon}\, \Big[\Psi^A  (x^\mu + \varepsilon n^\mu ) -  \Pi^A{}_B  (x^\mu + \varepsilon n^\mu, x^\mu ) \Psi^B (x)  \Big]  \nonumber \\
&=& n^\mu \Big[ \partial_\mu \Psi^A (x) + i  (A_\mu (x))^A{}_B \Psi^B (x)  \Big] \, .
\end{eqnarray}
Consequently,  the Fock-Ivanenko  covariant derivatives \cite{FI, F}  in  the fundamental representation,  with respect to  the  general connection $(A_\mu (x))^A{}_B $,   take the form
\begin{eqnarray}\label{eq:CodegAf}
({\cal D}_\mu)^A{}_B = \delta^A_B \partial_\mu + i  [A_\mu (x) ]^A{}_B  \, .
\end{eqnarray}

\subsubsection{Adjoint representation }

The adjoint representation is defined via the generators of the gauge algebra. Analogous to the previous case, the covariant derivative of a field $\Phi^A{}_B(x)$ in this representation is constructed from the difference between the field's value at a point $x^\mu + \varepsilon n^\mu$ and its parallel transport from $x^\mu$ to $x^\mu + \varepsilon n^\mu$. Using equations (\ref{eq:PaTr}) and (\ref{eq:PiABigi}) for an infinitesimal displacement, we obtain
\begin{eqnarray}\label{eq:Dcodg2}
n^\mu ({\cal D}_\mu \Phi)^A{}_B &=&  \lim_{\varepsilon \to 0} \frac{1}{\varepsilon}\,\Big[\Phi^A{}_B  (x^\mu + \varepsilon n^\mu ) - \Phi_\|^A{}_B (x^\mu + \varepsilon n^\mu ) \Big]  \nonumber \\
&=& \lim_{\varepsilon \to 0}  \frac{1}{\varepsilon}\,
\Big[\Phi^A{}_B  (x^\mu + \varepsilon n^\mu ) -  \Pi^A{}_C  (x^\mu + \varepsilon n^\mu, x^\mu ) \Phi^C{}_D (x) (\Pi^\dagger)^D{}_B  (x^\mu + \varepsilon n^\mu, x^\mu )   \Big]  \nonumber \\
&=& n^\mu \Big[ \partial_\mu \Phi^A{}_B  (x) + i [A_\mu (x), \Phi  (x) ]^A{}_B \Big] \, .
\end{eqnarray}

Consequently, the general covariant derivative in the adjoint representation takes the form
\begin{eqnarray}\label{eq:Dcodgf}
({\cal D}_\mu \Phi)^A{}_B =   \partial_\mu \Phi^A{}_B (x) + i [A_\mu (x), \Phi (x)]^A{}_B  \, .
\end{eqnarray}

\subsection{Universality of covariant derivative}

We now demonstrate that the covariant derivative satisfies the Leibniz rule for products of fields. This will be shown in both the fundamental and adjoint representations, for the products $\Psi_1^A \Psi_2^B$ and $(\Phi_1)^A{}_C (\Phi_2)^C{}_B$, respectively.

In the fundamental representation, the definition yields
\begin{eqnarray}\label{eq:Dcodg1}
n^\mu  [{\cal D}_\mu (\Psi_1 \Psi_2) ]^{A B} &=&  \lim_{\varepsilon \to 0}  \frac{1}{\varepsilon}\, \Big[(\Psi_1^A \Psi_2^B ) (x^\mu + \varepsilon n^\mu ) - (\Psi_{1 \|}^A \Psi_{2 \|}^B ) (x^\mu + \varepsilon n^\mu ) \Big]  \nonumber \\
&=& \lim_{\varepsilon \to 0}  \frac{1}{\varepsilon}\, \Big[(\Psi_1^A \Psi_2^B ) (x^\mu + \varepsilon n^\mu )
-  \Pi^A{}_C  (x^\mu + \varepsilon n^\mu, x^\mu ) \Psi_1^C (x)  \Pi^B{}_D  (x^\mu + \varepsilon n^\mu, x^\mu ) \Psi_2^D (x)  \Big]  \nonumber \\
&=& n^\mu \Big\{ \Big( \partial_\mu \Psi_1^A + i  (A_\mu)^A{}_C  \Psi_1^C \Big)   \Psi_2^B  +     \Psi_1^A  \Big( \partial_\mu \Psi_2^B + i  (A_\mu)^B{}_D  \Psi_2^D \Big)  \, .
\end{eqnarray}
Using (\ref{eq:CodegAf}), we thus find
\begin{eqnarray}\label{eq:Dcodg1f}
 [{\cal D}_\mu (\Psi_1 \Psi_2) ]^{A B} =  ({\cal D}_\mu \Psi_1)^A   \Psi_2^B  +  \Psi_1^A   {\cal D}_\mu \Psi_2^B \, .
\end{eqnarray}

In the adjoint representation, we have
\begin{eqnarray}\label{eq:Dcodg3}
n^\mu  [{\cal D}_\mu (\Phi_1 \Phi_2) ]^A{}_B &=&  \lim_{\varepsilon \to 0}  \frac{1}{\varepsilon}\, \Big[ (\Phi_1 \Phi_2 )^A{}_B (x^\mu + \varepsilon n^\mu )
- (\Phi_1 \Phi_2)_\|^A{}_B  (x^\mu + \varepsilon n^\mu ) \Big]  \nonumber \\
&=& \lim_{\varepsilon \to 0}  \frac{1}{\varepsilon}\, \Big[ ( \Phi_1 \Phi_2)^A{}_B  (x^\mu + \varepsilon n^\mu )
-  \Pi^A{}_C  (x^\mu + \varepsilon n^\mu, x^\mu ) (\Phi_1 \Phi_2)^C{}_D (x)   (\Pi^\dagger)^D{}_B  (x^\mu + \varepsilon n^\mu, x^\mu ) \Big]  \nonumber \\
&=& n^\mu \Big[ \Phi_1 \Big(  \partial_\mu  \Phi_2  + i [A_\mu, \Phi_2]  \Big)  + \Big(  \partial_\mu  \Phi_1  + i [A_\mu,\Phi_1]  \Big)  \Phi_2  \Big]^A{}_B    \, ,  \qquad
\end{eqnarray}
and  consequently
\begin{eqnarray}\label{eq:Dcodgf3}
 [{\cal D}_\mu (\Phi_1 \Phi_2) ]^A{}_B
=   \Big( {\cal D}_\mu  \Phi_1 \Big)^A{}_C  ( \Phi_2)^C{}_B     +  (\Phi_1)^A{}_C  \Big( {\cal D}_\mu  \Phi_2 \Big)^C{}_B   \, .
\end{eqnarray}
Therefore, the Leibniz rule holds in all cases.

As a concrete example, for spinor fields $\Psi^A \to \psi^\alpha$, the rule becomes
\begin{eqnarray}\label{eq:Dcodg1fe1}
 [{\cal D}_\mu (\psi_1 \psi_2) ]^{\alpha \beta} =    ({\cal D}_\mu \psi_1)^\alpha   \psi_2^\beta  +     \psi_1^\alpha  ( {\cal D}_\mu \psi_2)^\beta   \, .
\end{eqnarray}

Consider the vector   defined as
\begin{eqnarray}\label{eq:}
V^a =  {\bar \psi} \gamma^a \psi =  {\bar \psi}^\alpha \gamma^a_{\alpha \beta} \psi^\beta \,  .
\end{eqnarray}
Its covariant derivative is then
\begin{eqnarray}\label{eq:Vepbp}
({\cal D}_\mu   V)^a =  \gamma^a_{\alpha \beta} \Big[   ({\cal D}_\mu {\bar \psi})^\alpha   \psi^\beta  +     {\bar \psi}^\alpha   ({\cal D}_\mu
\psi)^\beta  \Big] \, .
\end{eqnarray}

Note that the covariant derivatives acting on the vector  and spinor  fields are defined with the generators appropriate to their respective representations.

\subsection{  Field strength  $ {\cal F}_{\mu \nu} (A)$  as  the commutator of covariant derivatives ${\cal D}_\mu$}

The replacement of ordinary derivatives with covariant derivatives has profound physical consequences. Firstly, it introduces a new field the connection  $A_\mu$ which describes interactions. Secondly, and in contrast to free theories, the  covariant derivatives  in an interacting theory are non-commutative.

To see this, we derive the commutator of two covariant derivatives. Applying   $( {\cal D}_\mu  {\cal D}_\nu )^A{}_B $ to a field    $\Psi^B$   in the fundamental representation yields
\begin{eqnarray}\label{eq:SpC}
[{\cal D}_\mu,  {\cal D}_\nu]^A{}_B  \Psi^B
=   i ( \partial_\mu A_\nu - \partial_\nu A_\mu)^A{}_B  \Psi^B  - [ A_\mu,    A_\nu ]^A{}_B      \Psi^B    \,  .
\end{eqnarray}

We now define the field strength tensor as
\begin{eqnarray}\label{eq:FSGe}
( {\cal F}_{\mu \nu} (A) )^A{}_B = \Big( \partial_\mu A_\nu - \partial_\nu A_\mu + i [ A_\mu,  A_\nu ] \Big)^A{}_B     \,   .
\end{eqnarray}
Using this definition, the commutator \eqref{eq:SpC} simplifies to
\begin{eqnarray}\label{eq:SpCf}
[{\cal D}_\mu,  {\cal D}_\nu]^A{}_B    = i ( {\cal F}_{\mu \nu} (A) )^A{}_B     \,  .
\end{eqnarray}
Note that $[{\cal D}_\mu,  {\cal D}_\nu]^A{}_B $  in this context is not a differential operator, but acts as a multiplicative factor.

\end{document}